\documentclass[useAMS]{mn2e} 
\usepackage{amssymb,amsmath}
\usepackage{graphicx} 
\usepackage{epstopdf}
\usepackage{journal_shortcuts}

\def\etal{{\it et al.\ }}  \def\eg{{\it e.g.\ }}  \def\msun{{M_\odot}}
\def\sun{{\odot}}

\newcommand{\be}{\begin{equation}}  \newcommand{\ba}{\begin{eqnarray}}
\newcommand{\ee}{\end{equation}}  \newcommand{\ea}{\end{eqnarray}}

\newcommand{\DM}{_{\mathrm{DM}}} \newcommand{\Gas}{_{\mathrm{gas}}}

\newcommand{\kpc}{\,\textrm{kpc}}

\title[Simulating Supersonic Turbulence in Galaxy Outflows]%
      {Simulating Supersonic Turbulence in Galaxy Outflows}

\author[E. Scannapieco \&  M. Br\"uggen]{Evan Scannapieco$^1$,
Marcus Br\"uggen$^2$ \\ $^1$School of Earth and Space Exploration,
Arizona State University, P.O.  Box 871404, Tempe, AZ, 85287-1404,
USA\\ $^2$Jacobs University Bremen, P.O. Box 750\,561, 28725 Bremen,
Germany\\ }

\begin{document}

\date{Accepted. Received; in original form }

\pagerange{\pageref{firstpage}--\pageref{lastpage}} \pubyear{2009}

\maketitle

\label{firstpage}

\begin{abstract}

We present three-dimensional, adaptive mesh simulations of dwarf
galaxy outflows driven by supersonic turbulence.  Here 
we develop a subgrid model to track not only the thermal and
bulk velocities of the gas, but also its turbulent velocities and
length scales.    This allows us to  deposit energy from supernovae
directly into supersonic turbulence, which acts on scales much larger
than a particle mean free path, but much smaller than resolved large-scale 
flows.  Unlike previous approaches, we are able to simulate a
starbursting galaxy modeled after NGC 1569, with realistic radiative
cooling throughout the simulation.   Pockets of hot, diffuse gas
around individual OB associations sweep up thick shells of material
that persist for long times due to the cooling instability.  The
overlapping of high-pressure, rarefied regions leads to a collective
central outflow that escapes the galaxy by  eating away at the
exterior gas through turbulent mixing, rather than gathering it into a
thin, unstable shell.  Supersonic, turbulent gas naturally avoids
dense regions where turbulence decays quickly and cooling times are
short, and this further enhances density contrasts throughout the
galaxy-- leading to a complex, chaotic distribution of bubbles, loops
and filaments as observed in NGC 1569 and other outflowing
starbursts. 

\end{abstract}

\begin{keywords}
galaxies:starburst--hydrodynamics--galaxies:dwarf
\end{keywords}

\section{Introduction}

Theoretical work has long shown that supernovae in low-mass galaxies
should  produce energetic  outflows that heat and enrich their
environments (Larson 1974; Dekel \& Silk 1986; Vader 1986).  Since
then such outflows have been observed in starbursting galaxies of all
masses (\eg Axon \& Taylor 1978; Heckman 1990; Bomans \etal 1997;
Martin 1998; 1999; Heckman \etal 2000; Schwartz \& Martin 2004;
Veilleux \etal 2005) and at all cosmological epochs (\eg Franx \etal
1997; Pettini \etal 1998; 2001; Frye, Broadhurst, \& Benitez 2002;
Rupke \etal 2005).  The existence of these ubiquitous galaxy outflows
has several important implications for galaxy formation.  The ejection
of heavy elements has been invoked to explain the strong correlation between
mass and metallicity observed in low-mass galaxies   (\eg Dekel \&
Silk 1986;  Richer \& McCall 1995; Mateo 1998;  Thacker \etal 2002;
Tremonti \etal 2004; Erb \etal 2006; Kewley \& Ellison 2008).  The
suppression of gas accretion onto starbursting galaxies (Benson \etal
2003) and onto neighbouring density perturbations (Scannapieco \etal
2000; Scannapieco \etal 2002) has been shown to be crucial to
reconcile the small number of observed dwarf galaxies with the large
number of low mass dark-matter halos in the favored cosmological model
(\eg Somerville \& Primack 1999; Cole \etal 2000; Benson \etal 2003;
Dekel \& Woo 2003).  The ratio of the baryonic mass to the gravitating
mass of galaxies has been found to be several times less than the
cosmic ratio (Hoekstra \etal 2005; Mandelbaum 2006), meaning that
either the baryons never fell into galaxies or that powerful galactic
winds removed them.  And, widespread galaxy outflows have proven to be
essential to understanding the history of the intergalactic medium
(IGM), which is observed to be widely enriched with heavy elements
(Tytler \etal 1995; Songaila \& Cowie 1996; Rauch \etal 1997; Chen
\etal 2001; Simcoe \etal 2002; Schaye \etal 2003; Aracil \etal 2004;
Adelberger \etal 2005; Scannapieco \etal 2006), that drastically
change its cooling properties (\eg Sutherland \& Dopita 1993; Wiersma
\etal 2008). 

It has also become clear that most of the outflows that play a key
role in each of these processes  are driven by core-collapse SN ejecta
and winds from massive stars.  These create hot, metal-enriched
bubbles that expand into the interstellar medium (ISM) and eventually
break out of the host galaxy in the form of bipolar outflows that can
reach velocities of hundreds of km/s (Veilleux \etal  2005). During this
process the bubbles sweep up cooler ambient gas that can also be blown
out of the galactic disk.   In fact, all current velocity measurements
made in starburst galaxies are of the entrained cooler material, which
is measured from UV or optical emission or via absorption lines (\eg
Rupke \etal 2005; Martin 2005). However, theoretical models predict
that around 90\% of the energy and metal content of the winds exist in
the hot ($T \geq 10^6$ K) phase, which has only recently been detected
in X-ray emission (Strickland \& Heckman 2007). 

The fact that the majority of the most important material remains invisible has
led to a wide range of assumptions regarding the efficiency with which
starburst-driven winds can eject metal-enriched gas.  Some authors
claim that only winds from dwarf galaxies can reach the IGM (Ferrara
\& Tolstoy 2000) while other claim that winds can also escape from
massive galaxies (Strickland \etal 2004).  Compounding this
observational uncertainty is the theoretical problem that the
starbursting disk itself is not well modeled as a single-temperature
medium in hydrostatic balance.  Rather the gas is both constantly
cooling and condensing into molecular clouds and being stirred and
heated by ionization fronts, stellar winds, and supernovae (\eg McKee
\& Ostriker 1977).  This combination of extremely short cooling times
and constant driving leads to a supersonic medium in which
the turbulent velocities exceed the thermal velocities, and turbulent
eddies act to support the disk even as they compress a fraction of
gas, driving star formation (\eg Mac Low \& Klessen 2004).
                                  
To approximate this configuration, previous simulations have been
forced to both artificially increase the mass-averaged ISM temperature
and suppresses its cooling through a variety of relatively crude
techniques.  The two-dimensional outflow simulations in  Mac Low \etal
(1989) and Mac Low \& Ferrara (1999), for example, included an
empirical heating function that was tuned to balance cooling in their
initial configuration, but  taken to be linearly proportional to the
density, such that it was overwhelmed by cooling within dense regions
that developed during the simulation.  Strickland \& Stevens (2000)
following Tomisaka \& Bregman (1993), Tenorio-Tagle \& Mu\~ noz-Tu\~
n\'on (1998), and Suchkov \etal (1994) imposed a minimum temperature
of $6.5\times 10^4$K in their two-dimensional simulations to account
for turbulent support of the disk.  D'Ercole \& Brighenti (1999)
imposed a temperature floor in their simulations,  equal to the $4.5
\times 10^3$K temperature of the initial ISM, and  Fujita \etal (2003;
2004) employed  a similar hard cutoff at $10^4$K.   Mori \etal (2002)
studied supernova feedback with full atomic cooling in a spherical
``pregalactic'' system with a $2 \times 10^4$K virial temperature that
was low-enough that catastrophic cooling of the inital was not a
problem. Finally Cooper \etal (2008) carried out simulations with an
inhomogeneous ISM model made up of dense clouds of $T \leq 3 \times
10^4 K$  material surrounded by much more tenuous $5 \times 10^6 K,$
``halo gas'' such that very little gas in their simulations was
located near the peak of the cooling function.

While the extremely short cooling times within the ISM make it
impossible to model supernova feedback by simply adding thermal energy
to the gas, at the same time, the range of physical scales involved 
does not allow for the direct simulation of supernova
remnants within a galaxy-scale simulation. Again, many
approximations have been made to avoid this problem: ranging from
temporarily lowering the densities of heated particles (Thacker \&
Couchman 2000), to delaying their cooling (Gerritsen \& Icke 1997), to
implementing momentum kicks rather than heating (Navarro \& White
1993; Mihos \& Hernquist 1994;  Scannapieco \etal 2001; Springel \&
Hernquist 2003).  These theoretical uncertainties have also lead to
suggestions that direct driving by supernovae and stellar winds (\eg
Silk 1997) may not be effective at all, but rather the primary driver
might involve additional physics such as radiation pressure on dust
(Thompson \etal 2005),  or non-thermal pressure caused by cosmic rays
(Socrates \etal 2008).  In fact some of the scaling relations from
such alternative models have been shown to reproduce many of the
observed trends seen in the intergalactic medium (Oppenheimer \&
Dav\'e 2006).

In this paper, we present simulations of starburst-driven outflows
from a dwarf galaxy using an entirely new approach.  Building on a
method developed by Dimonte \& Tipton (2006), we develop a subgrid
turbulence model that accounts both for the turbulent support of the
disk and the extra turbulent energy input  that drives a global
outflow.  Properly accounting for turbulence leads to a model for the
galaxy that has a realistic temperature and pressure distribution,
while at same time including cooling throughout the simulation.
Furthermore, adding supernova input as turbulent energy lets us
include this contribution without tracking supernova remnants directly
or adopting arbitrary approximations to avoid the energy from being
immediately  radiated away.  This then allows us to make observational
predictions for the structure of the highly-disturbed ISM as well as
assess the role of turbulent mixing on the escape fractions of
gas, kinetic energy, and metals.

The structure of the paper is as follows: In \S2 we describe our
galaxy model, feedback model, and subgrid model for supersonic
turbulence.  Whenever possible we  tune our parameters to approximate
the starbursting dwarf galaxy  NGC 1569, whose outflow has been
well-studied at a variety of wavelengths.    In \S3 we present the
results of our simulations, examine their dependencies on run
parameters, and assess their observational  consequences. Conclusions
are given in \S4.

\section{Method}

\subsection{Simulation and Model Galaxy}

All simulations were performed with  FLASH version 3.0, a
multidimensional adaptive mesh refinement hydrodynamics code  (Fryxell
\etal 2000) that solves the Riemann problem on a Cartesian grid using
a directionally-split  Piecewise-Parabolic Method (PPM) solver
(Colella \& Woodward 1984; Colella \& Glaz 1985; Fryxell, M\" uller,
\& Arnett 1989).  In all runs we simulated a galaxy  made up of a
gas+stellar disk,  contained within  a dark matter halo.  As a
prototypical dwarf starburst, we chose parameters to approximate the
nearby galaxy NGC 1569, which has been well observed and analyzed in a
wide variety of wavelengths (\eg Reakes 1980; Israel 1988; Gonz\' alez
Delgado \etal 1997; Martin 1998; Greggio \etal 1998;  Heckman \etal
2001;  Stil \& Israel 2002; Martin, Kobulnicky, \& Heckman 2002)
Consistent with observations of this and other starbursting galaxies,
we considered a  gas distribution with nearly exponential radial and
vertical profiles (\eg Barazza \etal 2006).  However, as in Roediger
\& Br\" uggen (2006) we softened the distribution in both directions,
in order to  prevent steep density gradients in the galactic plane and
centre.  This gave a gas distribution of  \be
\rho(r,z)=\frac{M\Gas}{2\pi a\Gas^2 b\Gas}  \frac{{\rm sech}}{2 \zeta}
\left(\frac{r}{a\Gas}\right)\,  \frac{{\rm sech}}{\pi/2}
\left(\frac{|z|}{b\Gas}\right), 
\label{eq:dens_exp_soft}
\ee where $(r,z)$ are the radius and distance from the plane in
galactic cylindrical coordinates, $M\Gas$ is the total gas mass, the
vertical scale lengths are $a\Gas$ and $b\Gas$, respectively, and
$\zeta = 0.9159$ is Catalan's constant.  For $r\gtrsim a\Gas$ and
$|z|\gtrsim b\Gas$, this density distribution converges towards the
usual exponential disk $\rho(r,z)=\frac{M\Gas}{2\pi a\Gas^2 b\Gas}
\exp(-r/a\Gas)\,\exp(-|z|/b\Gas)$, and  we fix  $M\Gas = 2 \times 10^8
M_\odot$ (Israel 1988), $a_{\rm gas}=0.7$ kpc, and $b_{\rm gas} = 0.2$
kpc (Reakes 1980).

We did not calculate the self-gravity of the gas explicitly, but
rather as in  Roediger \& Br\" uggen (2006) we  modeled  the
gravitational potential of the gas and stars as a Plummer-Kuzmin disk
(Miyamoto \& Nagai 1975; Binney \& Tremaine 1987), which approximates
the exponential distribution in disk galaxies, but is much more
manageable analytically.  The gravitational potential in this case is
\be \Phi_{\rm disk}(r) = -  \frac{G M_{\rm disk}} {\sqrt{r^2 + \left[
      a_{\rm disk}+(z^2+ b_{\rm disk}^2)^{1/2} \right]^2}}, \ee where
$a_{\rm disk}=a\Gas,$ $b_{\rm disk}=b\Gas,$  and $M_{\rm disk}=3
\times 10^8 M_\odot$ are the radial scale length  vertical scale
length, and total mass, respectively, and $G$ is the gravitational
constant.  We also added a second contribution to this potential  due
to the dark matter halo in which the galaxy is contained.  In this
case we assumed a Burkert (1995) model (see also Mori \& Burkert
2000), which is given by  
\ba \Phi_{\rm DM}(R)= &-& \pi G \rho_{\rm
  d0} r_0^2  \left\{- 2 \left(  1 + \frac{R}{R_0}      \right) \arctan
\frac{R}{R_0} \right. \nonumber  \\ & &   \,\,\, +2 \left(  1 +
\frac{R_0}{R} \right) \ln\left(  1 + \frac{R}{R_0}    \right)
\nonumber \\ & & \,\,\,-\left. \left(  1 - \frac{R_0}{R}     \right)
\ln \left[ 1 + \left(\frac{R}{R_0}\right)^2   \right]  + \pi \right\}, 
\label{eq:pot}
\ea  where $R$ is the radius from the centre of the galaxy in {\em
  spherical} coordinates, $R_0$ is the core radius of the halo, and
$\rho_{\rm d,0} = 3.08 \times 10^{-24} \,  {\rm g} \, {\rm cm}^{-3} \,
(R_0/{\rm kpc})^{-2/3}$  is the central density of the halo.   For
this potential, the maximum circular velocity of the halo as a function
of $R_0$ is   
\be
v_{\rm c,max}  = 23.1 \, {\rm km} \, {\rm s}^{-1} \left(\frac{R_0}{\rm
  kpc}\right)^{2/3},  \ee  and for our model dwarf we assumed $v_{\rm
  c, max} = 35$ km/s as in NGC 1569 (Martin 1998).

Outside of the disk, the gas was assumed to consist of a uniform
medium, with a mean density of $\rho_{\rm ambient} = 10^{-28}$  g
cm$^{-3}$, which is $\approx$ 200  times the mean $z=0$ cosmological
density.   This gas was taken to be non-rotating, and in hydrostatic
balance with the assumed gravitational potential asymptotically
approaching $T_{\rm ambient}=2 \times 10^{4} K $ at large distances.   
Finally, we assumed
that metals were smoothly distributed within the galaxy, such that all
cells within the galaxy were initially at a fixed metallicity, $Z_{\rm
  init} = 0.25 Z_\odot$ (Gonz\' alez Delgado \etal 1997).  For
simplicity, we defined the boundary of the galaxy at $\rho = 10
\rho_{\rm ambient}$, taking $Z=0.01 Z_\odot$ outside of this region.

All our simulations were performed in a three-dimensional  25 kpc
$\times$ 25 kpc $\times$ 30 kpc region, with outflow boundaries on all
sides.  For our grid, we chose a block size of $8^3$ zones and an
unrefined root grid with $10 \times 10 \times 12 $ blocks, for a
native resolution of $313$ pc.   The refinement criteria were the
standard density and pressure criteria, and for our fiducial runs we
allowed for 3 levels of refinement beyond the base grid, corresponding
to  an  effective resolution of $640 \times 640 \times 728$ zones, 39
parsecs on a side.

\begin{figure}
\centerline{\includegraphics[height=9cm]{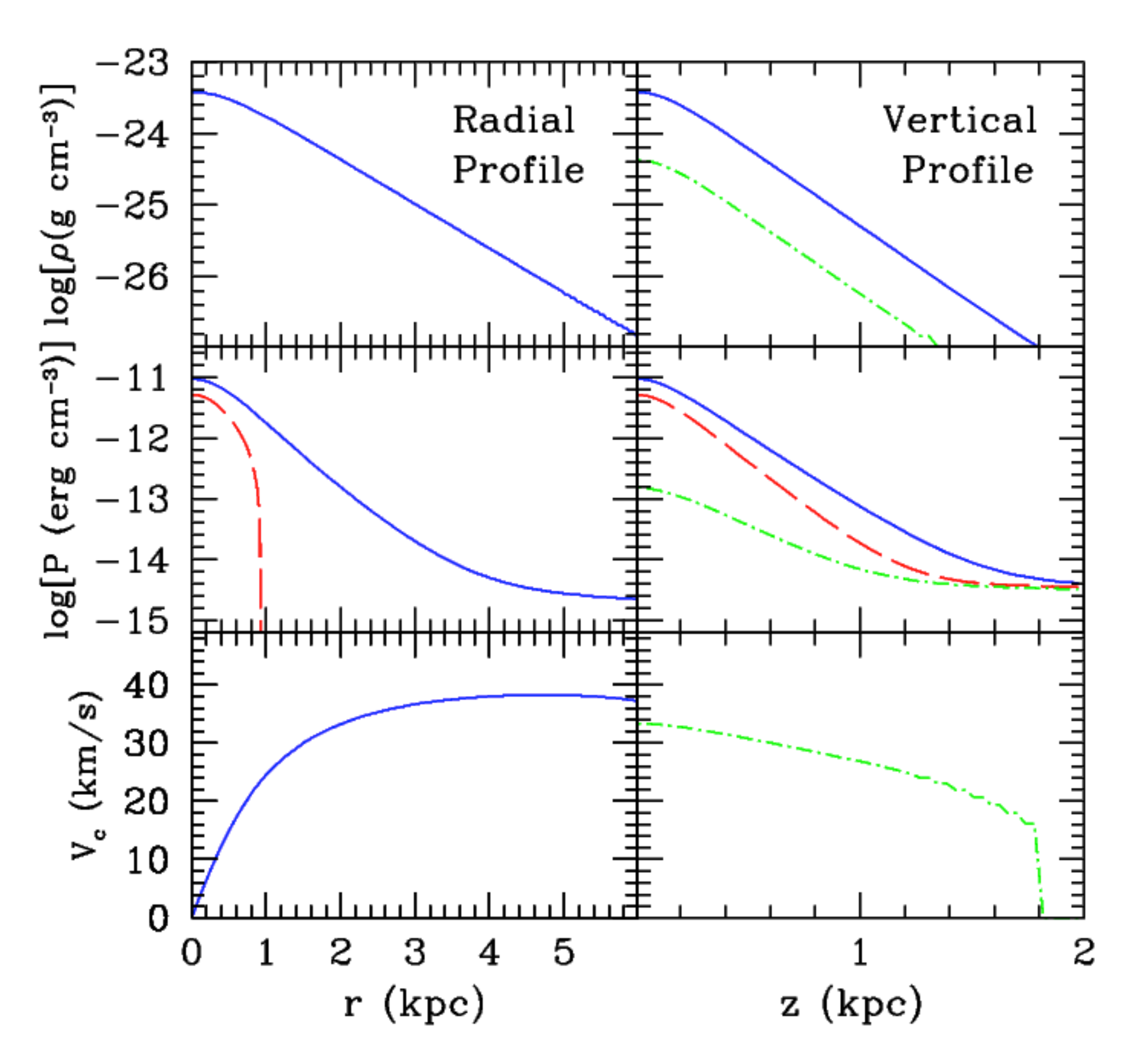}}
\caption{Initial set-up for our galaxy outflow simulations. {\em Top
    Left:} Density  in the midplane as a function of galactocentric
  radius.   {\em Top Right:} Density as a function of  height at $r=0$
  (solid line) and $r=2$ kpc (dot-dashed line).  {\em Centre left:}
  Total pressure as a function of radius in the midplane (solid line),
  and the component of pressure in the midplane arising from
  turbulence (dashed line).  {\em Centre right:} Total pressure as a
  function of height at $r=0$ (solid line), and $r=2$ (dot-dashed
  line) and the component of pressure at $r=0$ arising from
  turbulence (dashed line).  {\em Bottom left:} Circular velocity as a
  function of radius in the midplane.  {\em Bottom right:} Circular
  velocity as a function of height at $r=2$.}
\label{fig:initial_profiles}
\end{figure}

%
\begin{table}
\caption{Fixed model parameters.}
\label{tab:galaxy_parameters}
\centering\begin{tabular}{lll} \hline   
Component & Parameter & Value \\ 
\hline \hline    gas &$a\Gas$     & 0.7 kpc \\  
&$b\Gas$     & 0.2 kpc \\  
&$M\Gas$ & $2 \times 10^8 \msun$ \\  
& SFR         & 0.25 $M_\odot$/yr  \\  
&$Z\Gas$     & 0.25 $Z_\odot$ \\  
gravitational  & $a_{\rm disk}$   & 0.7 kpc  \\   
potential   & $b_{\rm disk}$   & 0.2 kpc  \\   
&$M_{\rm disk}$   & $3 \times 10^8 \msun$ \\   
DM halo &$r\DM$      & 2 kpc \\ 
&$v_c$     & 35  km/s \\  \hline
\end{tabular}
\end{table}
%

Having fixed the density distribution in the disk, its pressure,
turbulence, and temperature distribution were set such that: (i)
hydrostatic balance was maintained in the direction perpendicular to
the disk plane; and (ii) throughout the galaxy $T \leq 10^4$ K as
atomic cooling quickly radiates  thermal energy away above this
temperature.  In radial direction, the circular velocity was set so
that the centrifugal force balanced the gravitational force and
pressure gradients.  Figure~\ref{fig:initial_profiles} shows the
resulting radial and vertical profiles of density, pressure and
circular  velocity for our fiducial dwarf-starburst model.  The
choice of parameters for this model are summarized in Table
\ref{tab:galaxy_parameters}.

\subsection{Star Formation and Feedback}

For each galaxy, we computed a star-formation rate per unit area,
$\Sigma_{\rm SFR}$ from the surface density of gas, $\Sigma\Gas$
according  to the Kennicutt-Schmidt law  \be  \Sigma_{\rm SFR} =  2.5
\times 10^{-4} \, \frac{M_\sun}{{\rm yr} \kpc^2} \left(
\frac{\Sigma\Gas}{10^6 M_\sun {\rm kpc}^{-2}} \right)^{1.5},
\label{eq:KS}
\ee  which provides an accurate fit over $\approx 6$ orders of
magnitude in $\Sigma\Gas$ (Schmidt 1959; Kennicutt 1998).  In this
case the total star formation rate is given by 
\be SFR =  6.3 \times
10^{-5} \, \frac{M_\sun}{\rm yr} \left( \frac{M\Gas}{10^6 M_\sun}
\right)^{1.5}  \left( \frac{a\Gas }{\kpc} \right)^{-1}, 
\ee which we
assumed to last for 50 Myrs in all of our simulations.  For our choice
of parameters this gave an overall star-formation rate of 0.25
$M_\sun$ yr$^{-1}$, which is comparable to that observed in NGC 1569
(Israel 1998; Greggio \etal 1998).  This  corresponds to a total mass
of $1.3 \times 10^7 M_\odot$ stars formed over the  course of each
simulation.

Each starburst was accompanied by energy and metal input from
supernovae. One estimate of this contribution comes from comparing the
cosmic SN rate per comoving Mpc$^3$ as measured by Dahl\' en \etal
(2004) with the cosmic star formation rate density as measured by
Giavalisco \etal (2004), which gives a core-collapse supernova rate of
$(7.5 \pm 2.5) \times 10^{-3}$ SNe per solar mass of stars  formed
(Scannapieco \& Bildsten 2005).  An alternative estimate is to count
the number of $M \geq 8 M_\odot$ stars, by assuming  a Salpeter
initial mass function with upper  and lower mass cutoffs equal to 120
$M_\odot$ and $0.1 M_\odot$, respectively (Scannapieco \etal 2002)
which gives 1 SNe per 136 $M_\odot$ of stars formed.  We compromised
between these two values and assumed that 1 SNe was generated per 150
$\msun$ of stars formed, releasing an energy of $10^{51}$ ergs.
Furthermore, we assumed that a fraction $f_{\rm w}$ of this energy was
instantaneously deposited into the galaxy, such that during the
starburst the mechanical energy input was   
\be
 L_{\rm mech} = f_{\rm
  w} \, 2.2 \times 10^{41} \, {\rm ergs} \, {\rm s}^{-1}
\left(\frac{\rm SFR}{M_\odot {\rm yr}^{-1}} \right),  
 \label{eq:Lmech}
\ee   where in our fiducial models $f_{\rm w} = 0.7$.  As described in
more detail below, this energy was added as {\em turbulent kinetic
  energy}, rather than as gas thermal energy, vastly changing the rate
at which it was lost to cooling.  

As in a real galaxy, supernova energy was deposited into the gas
stochastically, approximating the patchy distribution of OB
associations within which  stars and SNe are formed.  In nearby
galaxies, the luminosity function of OB associations is
well-approximated by a power law of the form  \be \frac{d N_{\rm
    OB}}{dN} = A N^{-\beta},  \ee  where $N_{\rm OB}$ is the number of
OB associations containing $N$ many OB stars, and $\beta \approx 2$
(McKee \& Williams 1997; Oey \& Clarke 1997).  To approximate this
distribution in our simulations, we drew random numbers  $a_i \in
[0,1]$ such that for each forming OB association $i$, the number of OB
stars was given by \be N_i = [a_i N_{\rm max}^{(1-\beta)} + (1-a_i)
  N_{\rm min}^{1-\beta}]^{1/(1-\beta)}, \ee where we took $N_{\rm min}
= 20$ and $N_{\rm max} = 1000.$ By drawing two other random numbers
$b_i, c_i \in [0,1]$ we assigned each OB association a random
azimuthal angle given by $\phi_i = 2 \pi c_i$  and a random radial
position given by the transcendental equation $r_i = \tilde r_i a_{\rm
  gas}/1.5$ where $\tilde r_i = \ln[(1+ \tilde r_i)/(1-b_i)].$ All
associations were assumed to be centered around the midplane of the
galaxy, and  we paused for a time $150 N_i/SFR$ between bursts to
maintain the overall star formation rate of the simulation.   

For each OB association, we injected $f_{\rm w} 10^{51} N_i$ ergs of
turbulent kinetic energy into the simulation in a region of size
radius $r_{\rm bubble}$, which was at least the size of the region
containing twice the gas mass converted into stars, but no smaller
than 60 pc so as to avoid extremely high pressure, largely-unresolved
regions that excessively slow down the computational time step.  The
turbulent length scale for each OB association, discussed in more
detail below, was taken to be $r_{\rm bubble}.$ Each SN was also taken
to deposit additional gas and metals into this region.   Here we
assumed that the average ejected total mass of 8 $M_\odot$,  per SN, 2
$M_\odot$ of which is made up of heavy elements.   This is consistent
with the average stellar yields from a range of  SNe II simulations
(\eg Maeder 1992; Wooseley \& Weaver 1995; Arnett 1996; Tsujimoto
\etal 1995;  Nagataki 1998), although there are  significant
theoretical uncertainties between various estimates.

Metallicity-dependent radiative cooling was calculated in the
optically thin-limit, assuming local thermodynamic equilibrium:   \be
\dot E_{\rm cool}  =  
-(1-Y)(1-Y/2) \frac{\rho \Lambda(T,Z)}{(\mu   m_p)^2}(1+0.25 v^2_t/c_s^2),
\label{eq:ecool}
\ee where $\dot E_{\rm cool}$ is the radiated energy per unit mass,
$\rho$ is the density in the cell, $m_p$ is the proton mass, $Y$ is
the helium mass fraction, $\mu$ the mean atomic mass, and
$\Lambda(T,Z)$ is the cooling rate as a function of temperature and
metallicity, $V_t$ is the turbulent velocity, discussed below, and $c_s$ is
the local sound speed.  Here we made use of the tables compiled by  Wiersma,
Schaye, \& Smith (2009) from the code CLOUDY (Ferland \etal 1998),
making the simplifying approximation that the abundance ratios of the
metals both within the galaxy and ejected by the supernovae occured in
the solar proportions.  Furthermore, we also account for unresolved
substrucutre in highly turbulent regions, assuming that within each cell
the averaged density squared, $\left<  \rho^2 \right>$ is given by the bulk
density squared $\rho^2$ times an enchancment factor that increases with  
the Mach number of supersonic turbulence as $1 + 0.25 (v_t/c_s)^2$ as measured
by Padoan, Nordlund, \& Jones (1997).

Because the disk in our simulations was
initially supported primarily by turbulent pressure rather than
thermal pressure, and because the supernova energy was added to
turbulence rather than to the thermal motions, no approximate fixes to
this equation were necessary.  Instead, we were able to implement
cooling  in every cell in the simulation at every time step.

\subsection{Turbulence Modeling}

While the direct simulation of turbulence is extremely challenging,
computationally expensive, and dependent on resolution  (\eg Glimm
\etal 2001),  its behavior can be approximated to a good degree of
accuracy by adopting a subgrid approach. Recently, Dimonte \& Tipton
(2006) described a subgrid  model that is especially suited to
capturing the buoyancy-driven turbulent evolution of subsonic bubbles
(Scannapieco \& Br\" uggen 2008; Br\" uggen \& Scannapieco 2009),
which we have modified heavily to apply to supersonic galaxy-scale
outflows.  Other recent efforts at the subgrid modeling of 
turbulence on galactic and extragalactic scales are described in
Maier \etal (2009) and Shen \etal (2009).
                                   
Our model is based on the Navier-Stokes equations extended to include
a turbulent viscosity $\mu_T$  that depends on the eddy size
$L$ and kinetic energy per unit mass $K.$  The interaction between the
turbulence and the mean flow, is modeled by decomposing the flow into
average components and fluctuating components, for example ${\bf
  u}_{\rm tot} = {\bf u} + {\bf u}'$ the sum of the mass-averaged
mean-flow velocity and the fluctuating component of the velocity,
and ${\rho}_{\rm tot} = {\rho} + {\rho}'$ the sum of the 
mean-flow density and the fluctuating component of 
the density.  The
sum of these two components is   substituted back into Navier-Stokes
equations, which are then averaged to obtain separate evolutionary
equations for the mean and fluctuating components.  For compressible
flows, the averages are weighted by the density such that 
\be
\overline{ \rho_{\rm tot} {\bf u}'} \equiv 0, 
\label{eq:farve1}
\ee
and
\be
{\bf u}  \equiv  \frac{\overline{\rho_{\rm tot} {\bf u}_{\rm tot}}}{\rho}.
\label{eq:farve2}
\ee
The resultant equations constitute an expansion about the mean flow that must be terminated with a simplifying 
set of closure assumptions. 

To leading order, the mean flow fluid equations in this case are given by  
\ba
\frac{\partial {\rho}}{\partial t} + \frac{\partial {\rho}
  u_j}{\partial x_j} &=& 0,
\label{eq:rho}\\
\frac{\partial {\rho} u_i}{\partial t} + \frac{\partial {\rho} u_i
  u_j}{\partial x_j}  +\frac{\partial P}{\partial x_i} &=&  0,
\label{eq:u}\\
\frac{\partial {\rho} E}{\partial t} + \frac{\partial {\rho} E
  u_j}{\partial x_j}  +\frac{\partial P u_j}{\partial x_j}  &=&
\frac{\partial}{\partial x_j} \left( \frac{\mu_t}{N_E}  \frac{\partial
  E}{\partial x_j}\right)  \nonumber \\ 
& & \,\,\, + \rho \dot E_{\rm mech}  + \rho \dot E_{\rm cool}, 
\label{eq:E} 
\ea  
where $t$ and ${\bf x}$ are time and position variables, $
\rho({\bf x},t)$ is the average density field,  $u_i({\bf x},t)$  is
the mass-averaged mean-flow velocity field in the $i$ direction,
$P({\bf x},t)$ is the total pressure, both turbulent and kinetic,
 and $E({\bf x},t)$ is the mean
internal energy per unit mass, also including both turbulent and thermal
motions,  and $N_E=1$.   
Note that by modeling the impact of turbulence on the momentum equation by 
using on ly a pressure, we are neglecting off-diagonal terms  ($\partial u_i/\partial u_j + 
\partial u_j/\partial u_i,$ where $i \ne j$.)   This is because in the presence of shocks, 
such strain terms become unphysically large (\eg Gauthier \& Bonnet 1990; Klem 2004).  
Although these shortcomings may be overcome with appropriate use of limiters 
(\eg Wilcox 1994; Sinha \etal 2003), we set this aside as a future refinement to the basic
method presented here.

In eq.\ (\ref{eq:E}),
the first term on the right
captures the effects of  turbulent
mixing, which is modeled as a turbulent viscosity $\mu_t.$ 
the second, $\dot E_{\rm mech}$ term is an explicit source term that is
determined by the mechanical luminosity as given by
eq.\ (\ref{eq:Lmech})  and the size of the region into which the
energy is being added, and the third, $\dot E_{\rm cool},$ term is given by
eq.\ (\ref{eq:ecool}).  
Note that $\mu_t$ does not appear in the continuity equation
due to eq.\  (\ref{eq:farve1}) and does not appear in the momentum equation due to 
eq.\ (\ref{eq:farve2}). 

In any type
of subgrid model a number of fit parameters such as $N_E$ arise, whose
values are expected to be $\approx 1,$ but must ultimately be fine
tuned versus experiments to achieve the most accurate results.  Here
we take the same fit parameters as used in Dimonte \& Tipton (2006),
and in Table 2 we summarize how they have been determined.
However, there are several important differences in our approach in this paper, 
and thus it is likely our model will eventually be able to be further
improved by re-adjusting these values to experiments and observations.
Unlike in Scannapieco \& Br\"uggen (2009),
$E$ is now the total internal energy, including both turbulent and
thermal contributions.   Likewise the pressure is the sum of both
thermal and turbulent component, computed as $P = \frac{2 \rho}{3 \mu
  m_p} E,$ with $m_p$ the mass of the proton and $\mu$ the mean atomic
weight of the gas.  This redefinition allows us to apply the PPM solver
to capture both the effects of turbulent and thermal pressures,
yielding accurate solutions even in cases in which most of the internal
energy is in the subgrid turbulent flow.  An implicit and simplifying
assumption  associated with this approach is that turbulent and thermal 
velocities provide pressure support through the same equation of state.
To track the metals ejected by supernovae,
eqs.\ (\ref{eq:rho})-(\ref{eq:E}) are supplemented by a mass-fraction
equation:  
\be  
\frac{\partial {\rho F_r}}{\partial t} +
\frac{\partial {\rho F_r} u_j}{\partial x_j}= \frac{\partial}{\partial
  x_j} \left(  \frac{\mu_t}{N_F} \frac{\partial F_r}{\partial
  x_j}\right)
\label{eq:massfrac},
\ee  where $F_r$ is the mass fraction of species $r$ in a given zone,
and $N_F =  1.0$ is a scale factor.

The turbulence quantities that appear in these equations are
calculated from evolution equations for $L,$ the scale of the
largest turbulent eddies, 
and $K,$  the turbulent kinetic energy.   Simple equations for the
evolution of these quantities are given by 
\be  
\frac{\partial \rho  L}{\partial t} + \frac{\partial \rho L u_j}{\partial x_j} =
\frac{\partial}{\partial x_j} \left(  \frac{\mu_t}{N_L}
\frac{\partial L}{\partial x_j}\right) + C_C \rho L \frac{\partial
  u_i}{\partial x_i}
\label{eq:L},
\ee  
and  
\ba  
\frac{\partial \rho K}{\partial t} + \frac{\partial 
  \rho K u_j}{\partial x_j} = &\frac{\partial}{\partial x_j} \left(
\frac{\mu_t}{N_K}  \frac{\partial K}{\partial x_j}\right) + \rho \dot
E_{\rm mech}  \qquad \qquad \\  
&  - \frac{K}{E} \frac{\partial P u_j}{\partial x_j}  
- \rho  V_t C_D \frac{\max(V_t-V_{t,0},0)^2}{L},   \nonumber
\label{eq:K}
\ea  
where $V_t \equiv \sqrt{2 K}$ is the average turbulent
velocity,  $V_{t,0}$ is the average turbulent velocity at the  start for
the simulation,  and   $C_C=1/3$ is given by mass conservation,  
$N_L=0.5$,   $N_K=1.0$, and
$C_D=1.25$ are experimental fit constants.  In the $L$ equation the
two terms on the right hand side  represent, respectively: 
turbulent diffusion and
the  growth of turbulent motions due to the expansion of the mean
flow.  In the $K$ equation the three terms on the right hand side
represent, respectively: turbulent diffusion,  the energy input from
SNe, and a term that causes turbulence to decay away at a
characteristic time scale of  $\approx L/(V_t-V_{t,0})$ in the absence
of external driving.   This energy goes directly into gas heating,
because of energy conservation $E = K + E_{\rm thermal}.$
The $V_t-V_{t,0}$ in this term means that  after the
starburst episode, the gas remaining in the galaxy will  eventually
settle back into a turbulently-supported disk, rather than dissipate
all its turbulence into heat.    Finally, the turbulent viscosity was
calculated as   
\be  
\mu_t =  \rho L \max(V_t-V_{t,0},0),
\label{eq:mut}
\ee  where again the
$V_t-V_{t,0}$ term assures that in its initial configuration the galaxy
remains static and does not diffuse into the intergalactic medium. 
Note that including the $V_{t,0}$ term in eqs.\ (\ref{eq:K}) and
(\ref{eq:mut}) is akin to asumming a low level of persistent turbulence
in addition to the much larger contribution by the SNe that are added to
the galaxy explicitly during the simulation.

\begin{table}
\caption{Summary of coefficients in the turbulence model.   Constants
that are fit to experiment appear with error bars, and in those cases
we take the central value for this study.}
\label{tab:ICM}
\centering
\begin{tabular}{llll} 
\hline 
Parameter & Value & Effect & Source\\
\hline 
$N_L$ & $0.5 \pm 0.1$ & Diffusion of $L$ & Experimental Fit\\ 
$N_E$ & $1.0 \pm 0.2$ & Diffusion of $E$ & Self-similarity\\
$N_F$ & $1.0 \pm 0.2$ & Diffusion of Species & Self-similarity\\
$N_K$ & $1.0 \pm 0.2$ & Diffusion of $K$ & Self-similarity \\ 
$C_C$ & $1/3$    &  Compression of $L$ & Mass Conservation\\ 
$C_D$ & $1.25 \pm 0.4$  & Drag term for $K$ & Experimental Fit\\ 
\label{table:DT}
\end{tabular}
\end{table}

Note that unlike in Dimonte \& Tipton (2006) and our previous
modeling, we do not include terms that attempt to track the growth of
the Rayleigh-Taylor or Richtmyer-Meshkov instabilities on subgrid
scales.  This is because we are no longer working in the regime   in
which turbulence generated by these instabilities is dominant.
Rather, we are interested in approximating the evolution of the ejecta
associated with a random collection of SNe that move supersonically
within overpressured bubbles surrounding OB associations.   Our
approach assumes that for each OB association, the initial maximum
turbulent length scale is approximately equal to the radius of the
overpressured region, and the initial turbulent kinetic energy per
unit mass is approximately equal to mechanical energy input from SNe
divided by the total mass in the region.  Although such overpressured
regions will eventually expand and become Rayleigh-Taylor unstable,
the length scale and velocities generated by this instability will
initially be much smaller than the motions within the bubble.  This
means that the turbulent regions will not grow according to the linear
equations of growth of the Rayleigh-Taylor or Richtmyer-Meshkov
instabilities.  In particular, one would not expect $L$ and $K$ in
eqs.\ (\ref{eq:L}) and (\ref{eq:K}), which represent SNe driven random
motions, to grow as quickly as $\propto t^2$ as they would in the
Rayleigh-Taylor case.  So our approach is to conservatively assume
that $L$ expands along with the mean flow, and that $K$ is driven only
by SNe input, and decays to thermal energy on the $L/V_{t}$ timescale
expected from a self-similar scaling analysis of incompressible turbulence
(Kolmogorov 1941),  which has also been confirmed in the compressible
(magnetohydrodynamic) case 
(Stone \etal 1998; Mac Low \etal 1998; Padoan \& Nordlund 1999).
Thus only {\it resolved} Rayleigh-Taylor and Richtmyer-Meshkov
instabilities are captured in our simulations, while
eqs.\ (\ref{eq:L}) and (\ref{eq:K}) track the much more important
subgrid turbulent motions driven by SNe.
		
As in our previous modeling, our numerical implementation of these
equations was divided into three steps, which were carried out after
the main hydro step in FLASH3, which advects all the variables above.
In the first step, we implement the $\partial u_i/ \partial x_i$ terms
in eq.\ (\ref{eq:L}) explicitly.    In the second step, we: (i)
compute $V_t$ as $\sqrt{2 K}$, (ii) use a leapfrog technique to add the
source term to $V_t$ as $S_K/\rho V$, and then (iii) write $V$ back to
the $K$ array as $K = V_t^2/2.$  Finally, in the third step, we
calculate the turbulent viscosity and use this to implement the
diffusive mixing terms in eqs.\ (\ref{eq:E})-(\ref{eq:K}) explicitly.
This final step requires us to impose an additional constraint on the
minimum times step of  $dt \leq (\Delta^2 \rho/\mu_t)/4$ where $(\Delta)$
is the minimum of  $dx$, $dy$, and $dz$ in any given zone.  This
diffusive constraint must be satisfied for all zones in the
simulation, but as discussed in SB08 this explicit approach works well
in concert with the AMR hydrodynamic solver,  because as $\mu_t$ increases
near the interface, density and pressure fluctuations are smoothed,
allowing the code to derefine in these regions. Thus the diffusive
time step remains greater than or comparable to the one required by
the Courant condition for most of the evolution in our
simulations.

In all zones within the galaxy, $K$ is initialized such that the
galaxy is in hydrostatic balance, even though the initial thermal
energy is chosen such that $T \leq 10^4$ everywhere.   Outside the
galaxy, the internal  energy was taken to be solely thermal such that
$K$ was initialized to be to $10^{-10} E.$ Throughout the simulation,
$L$ was initialized to 2 parsec, $1/100th$ of the initial gas scale height 
of the galaxy.


\section{Results}

\subsection{Fiducial Model}

\begin{figure*}
\centerline{\includegraphics[width=6.5in]{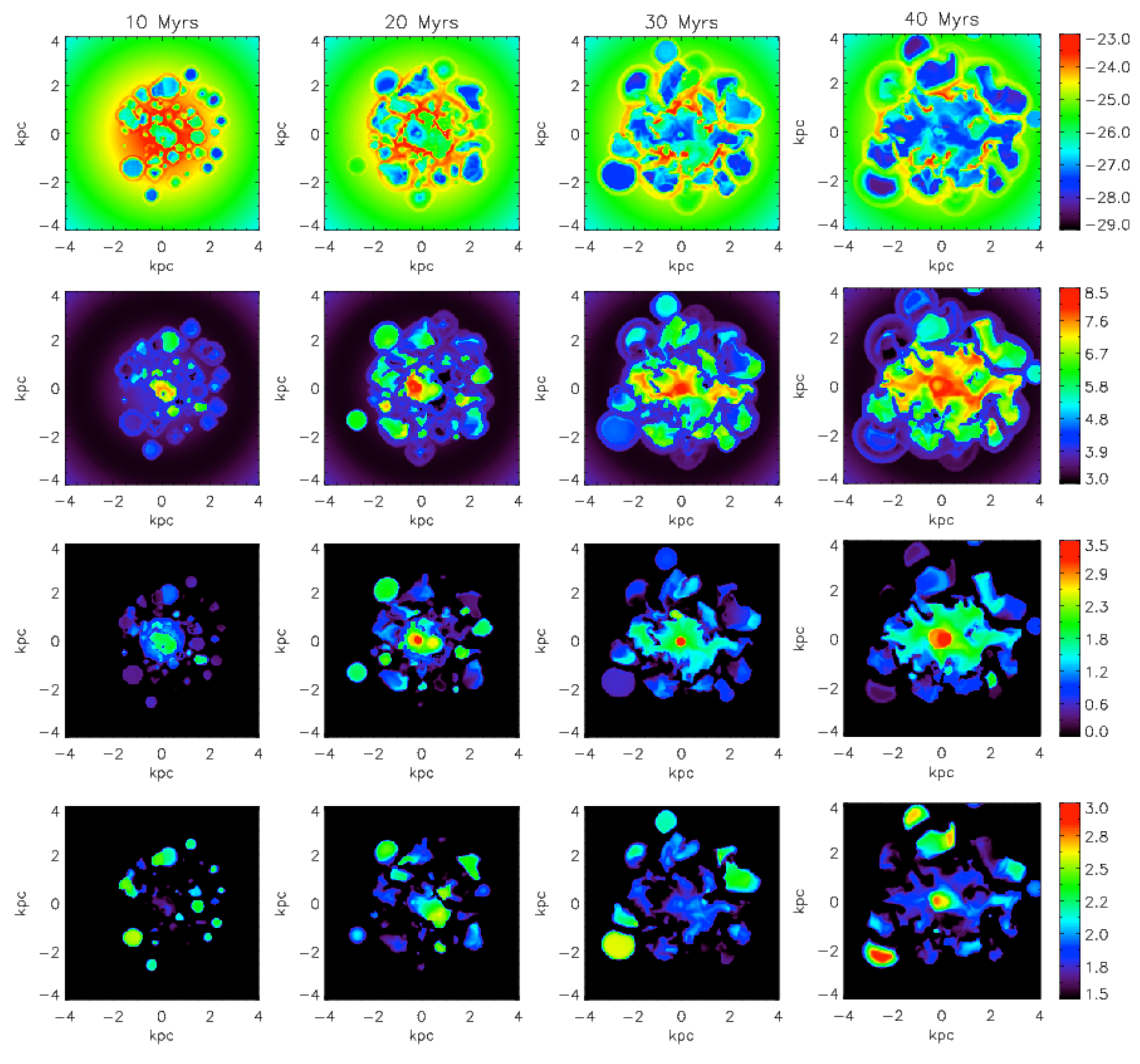}}
\caption{Horizontal slices through the central $8 \times 8$ kpc in our
  fiducial simulation (7-4D) at four representative times: 10, 20,
  30, \& 40 Myrs (arranged from left to right in each row).  {\em Top
    Row:} Contours of $\log$ density from $\rho = 10^{-29}$ to
  $10^{-23}$ g cm$^{-3}$.  {\em Second Row:} Contours of $\log$
  temperature from $T = 10^{3}$ to $10^{8.5}$ K.  {\em Third Row:}
  Contours of $\log$ turbulent velocity, $V_t$, from 1 to 3000 km
  s$^{-1}$. {\em Bottom Row:} Contours of $\log$ turbulent length scale, $L$ from 30 to 1000 parsecs.}
\label{fig:fiducial_zy_slice}
\end{figure*}

\begin{table}
\caption{Run Parameters}
\label{tab:runs}
\centering\begin{tabular}{lllll} \\ 
\hline  Run    & $f_{\rm w}$ &
Resolution  & Subgrid  & Clumpy \\ 
Name   &             &  (pc)   & & \\ 
\hline  7-4D      & 0.7  & 39  & Yes & No  \\ 
4-4D       & 0.4 & 39  & Yes & No  \\ 
2.5-4D       & 0.25  & 39  & Yes & No  \\ 
4-4N & 0.4  & 39  & No  & No   \\ 
7-3D      & 0.7  & 78  & Yes & No \\ 
7-2D      & 0.7  & 156  & Yes & No   \\ 
7-4DC     & 0.7  & 39  & Yes & Yes  \\
\hline
\end{tabular}
\end{table}

\begin{figure*}
\centerline{\includegraphics[width=6.5in]{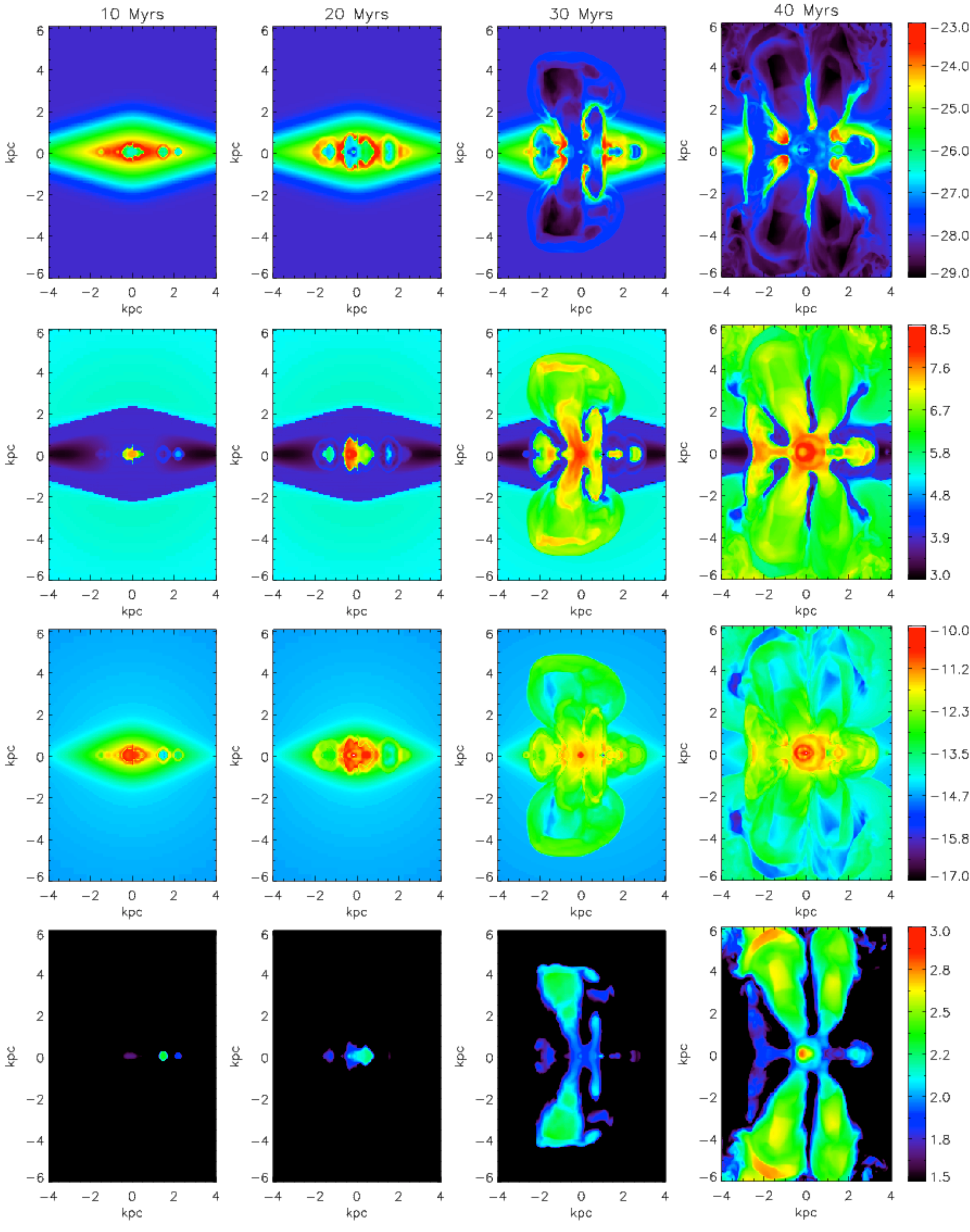}}
\caption{Vertical slices through the central $8 \times 12$ kpc in our
  fiducial simulation (7-4D) at four representative times of 10, 20,
  30, \& 40 Myrs (arranged from left to right in each row).  {\em Top
    Row:} Contours of $\log$ density from $\rho = 10^{-29}$ to
  $10^{-23}$ g cm$^{-3}$.   {\em Second Row:} Contours of $\log$
  temperature from $T = 10^{3}$ to $10^{8.5}$ K.  {\em Third Row:}
  Contours of $\log$ pressure from $p = 10^{-17}$ ergs cm$^{-3}$ to
  $10^{-10}$ ergs cm$^{-3}$. {\em Bottom Row:} 
Contours of $\log$ turbulent length scale, $L$ from 30 to 1000 parsecs.}
\label{fig:fiducial_xy_slice}
\end{figure*}

\begin{figure*}
\centerline{\includegraphics[width=6.5in]{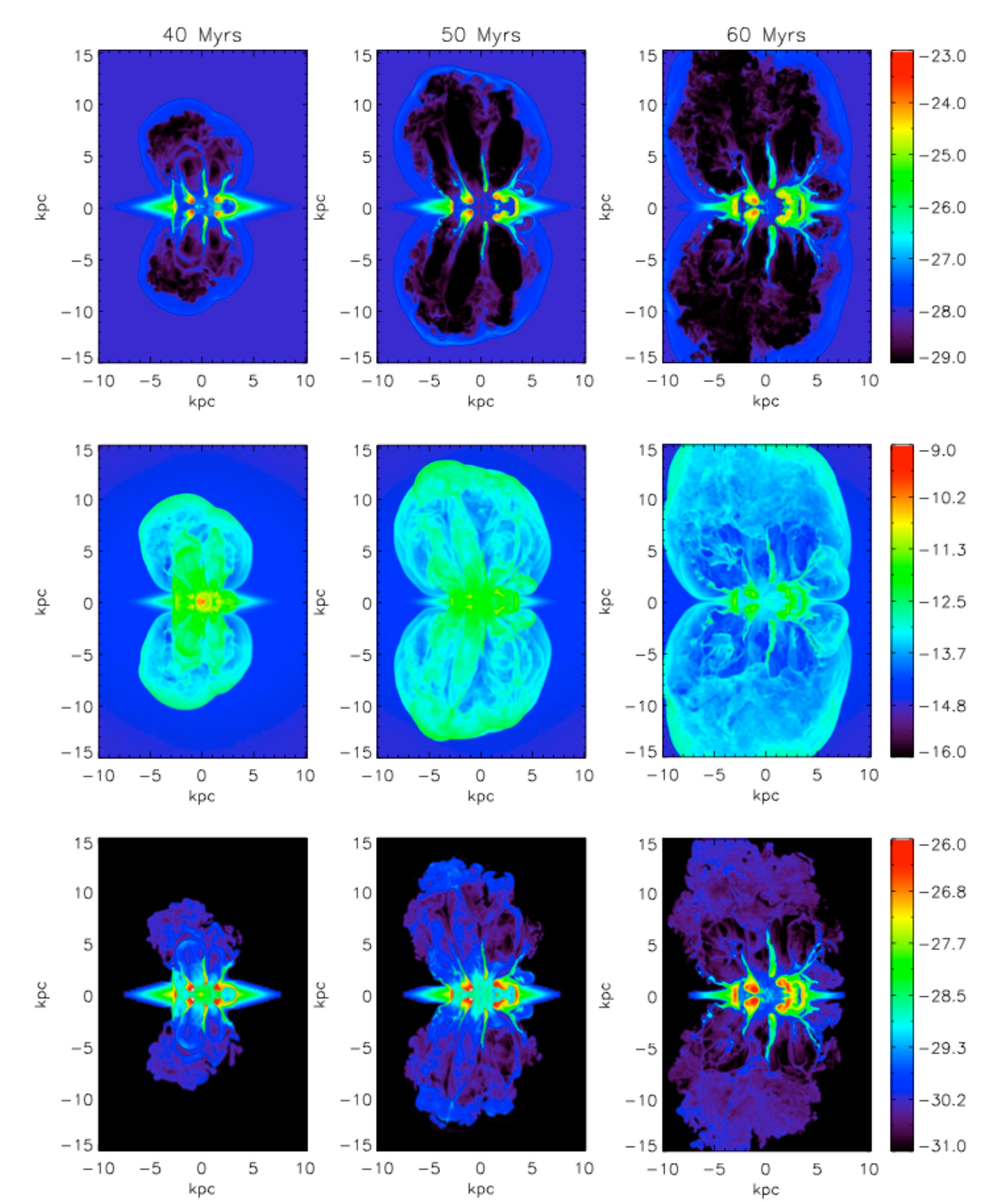}}
\caption{Vertical slices through the central $20 \times 30$ kpc in our
  fiducial simulations  at three representative times of 40, 50, \&
  60 Myrs (arranged from left to right in each row).  {\em Top Row:}
  Contours of $\log \rho$ from $\rho = 10^{-29}$ to $10^{-23}$ g
  cm$^{-3}$.   {\em Centre Row:} Contours of total $\log$ energy
  density (kinetic+internal) from $e_{\rm tot} = 10^{-16}$ to
  $10^{-9}$ ergs cm$^{-3}$.  {\em Bottom Row:} Contours of $\log$
  total metal density from $\rho_{\rm metals} = 10^{-31}$ to
  $10^{-26}$ g cm$^{-3}$.}
\label{fig:fiducial_xy_slice_large}
\end{figure*}

In Figs.\ \ref{fig:fiducial_zy_slice} -
\ref{fig:fiducial_xy_slice_large}  we show results from our fiducial
run, which has $f_{\rm w} =0.7$ and  4 total refinement levels,
including the base grid.    A summary of all the runs carried out for
this study is given in Table \ref{tab:runs}. Each run is labelled m-nD for
cases with subgrid diffusive mixing and m-nN for cases without
subgrid mixing, where m $= 10 \times f_{\rm w}$ and n is the number of
refinement levels.  Thus our fiducial run is referred to as 7-4D.

Fig.\ \ref{fig:fiducial_zy_slice} shows horizontal slices through the
central $8 \times 8$ kpc for this run at four representative times,
and Fig.\ \ref{fig:fiducial_xy_slice} shows 
vertical slices through the central $8 \times 12$ kpc.
These plots contain numerous
``superbubbles'' driven by SNe from individual OB associations,
as have been studied in several classic theoretical papers (Weaver \etal
1977; McCray \& Snow 1979; Tomisaka \& Ikeuchi 1986; Mac Low \& McCray
1988; Tenorio-Tagle \& Bodenheimer 1988; Mac Low \etal 1989).  
Note that the
bubbles are somewhat larger at greater distances from the galactic
centre as the ambient pressure is lower there.
Regardless of galactocentric distance, however, the bubbles are
significantly less empty than those described in the 
papers mentioned above.  This primarily because we assume that  SNe form
and deposit energy into a region containing twice as much gas mass as
is converted into stars. This means that for every $10^{51}$
ergs added in SN driven turbulent energy, the affected region
initially contains at least $2 \times 150 \msun$ of gas.  In
this case, the increase in internal energy per unit mass is $\approx
f_{\rm w} 500$ times that of the gas within the midplane, which
corresponds to an initial turbulent velocity of $\approx f_{\rm  w}^{-1/2} 500$ km/s.   

Within these regions three time scales are almost equal: (i) the time
scale for turbulent energy to dissipate into thermal energy, $t_{\rm
  diss} \approx r_{\rm bubble}/V_t;$ (ii) the time scale for bubble
expansion, $t_{\rm dyn} \approx r_{\rm bubble}/c_{\rm s,eff} \approx
r_{\rm bubble}/V_t,$  where the effective sound speed within the
heated regions is dominated by the turbulent motions, and (iii) the
time scale for mixing of the bubble interior with the surrounding
matter, $t_{\rm mix} \approx r_{\rm bubble}^2 \rho/\mu_t  \approx r_{\rm
  bubble}^2/(V_t r_{\rm bubble}) \approx r_{\rm bubble}/V_t.$
Physically, this similarity of time scales occurs because the driving
of gas   into the exterior medium and the mixing of the exterior
medium with the bubble interior both occur roughly at the time at
which SNe kinetic energy is converted into thermal energy
through a turblent cascade.
 Furthermore, all these time scales are short, on the order of  10
pc / $(f_{\rm w}^{-1/2} 500 \, {\rm km} \, {\rm s}^{-1})$ $\approx$ 0.2
Myrs, which is much less than their interior cooling time and the
dynamical time of the surrounding medium. 

Thus the bubbles quickly expand to the point that they are in pressure
equilibrium with their surroundings, with only moderate heating of
their interiors through turbulent decay, and moderate mixing of shells
with the interior gas through turbulent ``diffusion.''   
As $p \propto
n^{5/3} \propto l^{-1/5}$ during this approximately adiabatic expansion, the
result is the formation of $\approx (f_{\rm w} 500)^{1/5} \times 60 \,
{\rm pc} \approx f_{\rm w}^{1/5} 200$ pc superbubbles whose exterior
shells are relatively thick, and whose interiors are roughly  $(f_{\rm
  w} 100)^{3/5} \approx f_{\rm w}^{3/5} 40$ times underdense and
somewhat hotter and more turbulent than their surroundings.   
As the turbulent length scale expands along with the flow, $L$
also rises to $\approx 200$ pc within these regions, but remains well below
the grid scale throughout the rest of the simulation.  Many
such regions can be seen in Figs.~\ref{fig:fiducial_zy_slice} and
\ref{fig:fiducial_xy_slice}, particularly at later times and larger
galactocentric radii.  Note however, that turbulence
will tend to be strongest inside of the bubbles and 
outside of the swept up regions, as the  time scale
for turbulence  to decay to thermal energy is much smaller at high
densities.  This can be seen from eqs.\ (\ref{eq:L}) and (\ref{eq:K}),
which show that as $\rho L$ and $\rho K$ diffuse into denser regions
the time scale for turbulent dissipation drops as $t_{\rm decay}
\propto L/\sqrt{2 K} \propto \rho^{-1}/\sqrt{\rho^{-1}} \propto
\rho^{-1/2}.$  

\begin{figure*}
\centerline{\includegraphics[width=6.5in]{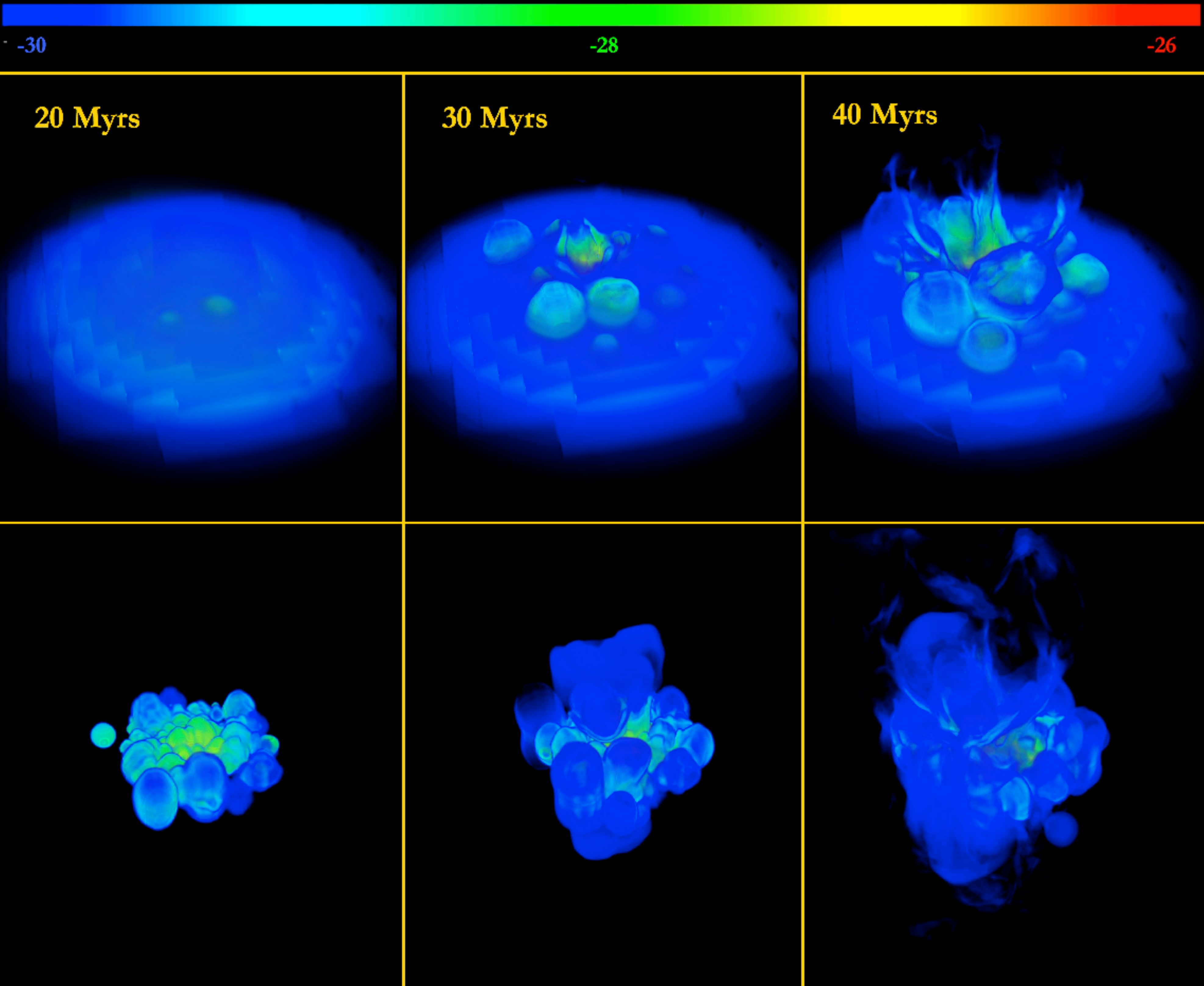}}
\caption{Rendered images of log density of metals from $10^{-30}$ to
$10^{-26}$ g cm$^{-3}$, separated into metals from the IGM (upper
  panels) and from the SNe that go off during the simulations (lower
  panels), at times of 20, 30, and 40 Myrs.  The galaxy is being
  observed at an inclination of $\approx 30$ degrees, and the galaxy
  disk visible in the upper panels has a radius of $\approx 6$ kpc.
  Metals from SNe driving the outflow are ejected to large distances
  while the ISM is largely  retained by the galaxy.   
 The plotting artifacts at large radii in the
 top panels  illustrate the $8^3$ cell ``blocks'' in our adaptive mesh approach,
 which has a resolution of 39 parsecs in the central regions where 
 superbubbles develop, but only 312 parsec at large radii, 
 where density and pressure contrasts are small.}
\label{fig:fiducial_metals_rendered}
\end{figure*}

Furthermore the fact that atomic cooling is implemented throughout the
simulation means the medium is unstable to the cooling instability,
and the dense shells around the bubbles will persist and grow over
time.   As described by Fall and Rees  (1985) in the absence of
thermal conduction and turbulence, dense clouds  will become cooler and
more compact over time if   
 \be
\frac{\Lambda(T_p,Z)}{T_p^{2}} > \frac{\Lambda(T_{\rm ISM},Z)}{T_{\rm
    ISM}^{2}},
\label{eq:cinstability}
\ee  where $T_p$ is the temperature of the perturbation, $T_{\rm ISM}$
is the temperature of the outside medium, and $\Lambda(T,Z)$ is the
cooling function.  In the case of atomic  cooling with $Z \approx
0.1 Z_\odot$, this is true whenever $T_p < T_{\rm ISM}$ and $T \gtrsim 3
\times 10^5 K,$ which is clearly the case in the hot rarefied medium
that develops within the galaxy.

 In our simulations turbulent pressure is also included, and
 eq.\ (\ref{eq:cinstability})
   is modified slightly to include the additional
 contribution to $p \, dV$ work,  yielding 
 \be
 \frac{\Lambda(T_p,Z)}{T_p^{2} (1+K_p/E_p)^2 } >
 \frac{\Lambda(T_{ISM},Z)}{T_{\rm ISM}^{2} (1+K_{\rm ISM}/E_{\rm
     ISM})^2 },
\label{eq:tcinstability}
\ee 
where $K_p/E_p$ is the fraction the internal energy in turbulence
within the perturbation, and $K_{\rm ISM}/E_{\rm ISM}$ is the fraction of
internal energy in turbulence in the exterior medium.  As turbulence
tends to persist longer in more rarefied media,
eq.\ (\ref{eq:tcinstability}) is even more easily satisfied than
eq.\ (\ref{eq:cinstability}).  This means that unlike in simulations
without cooling, the primary source of structure in our simulations is
not the Rayleigh-Taylor instability that leads to the fragmentation of
an initially smooth  shell, but rather  the
stochastic nature of SN heating, enhanced  by the cooling instability.
Thus gas condenses into denser structures long after pressure
equilibrium has been achieved between the swept up gas and the
exterior medium, as can be seen by comparing $\rho$ and $T$ to the
pressure as in Figure \ref{fig:fiducial_xy_slice}. Note
that a similar evolution of structure was also seen by  Mori \etal (2002),
who implemented cooling  throughout their simulations of small ``pregalactic
systems'' and by 
Cooper \etal (2008), who implemented cooling in a larger galaxy, modeled as
a hot rarefied medium surrounding distribution of cold, compact clouds.

Fig.\ 2 also shows a large diffuse region that develops  over
time near the centre of the galaxy.  Here the superbubbles begin to
overlap as star formation is strongly centrally concentrated due to
both the overall radial profile of the gas and the $\Sigma^{1.5}$
dependence of the Kennicutt-Schmidt law. Together, these overlaping outbursts
open a rarefied region that expands gradually over time.
As it grows, the collective outflow eats away at the exterior gas
through turbulent mixing, rather than gathering it into a thin,
fragile shell.  As a result, the rarefied region drills its way
almost directly vertically, following the path along which
the minimum amount of material separates the bubble interior from the
intergalactic medium.  ``Blow-out'' occurs when the overpressured
region resulting from overlapping OB associations moves into the
intergalactic medium, rather than when  the material surrounding a
single superbubble becomes Rayleigh-Taylor unstable.  This occurs
roughly at the time of bubble overlap, which can be estimated as 
\be
 t_{\rm overlap}(R) \approx A_{\rm bubble}(R)^{-1} M_{\rm
  OB} \Sigma^{-1}_{\rm SFR}(R), 
\ee 
where $M_{\rm OB}$ is the mass of a
typical OB association and $A_{\rm bubble}  \approx  2 M_{\rm OB}
\Sigma_{\rm gas}^{-1} (f_{\rm w} 500)^{2/5}$ is the area of the disk
covered by a bubble
after it expands to reach pressure equilibrium with its surroundings.
From eqs.\ (\ref{eq:KS})  and (\ref{eq:dens_exp_soft}) this gives 
\be
t_{\rm overlap}(R) \approx 15 \, {\rm Myrs} \, f_{\rm w}^{-2/5}  \exp\left(-\frac{r}{1.4 {\rm kpc}}\right) .
\label{eq:overlap}
\ee 
This means that for any value of the wind efficiency, bubble
overlap will occur much more quickly near the centre of the galaxy.
The result is a strongly bipolar outflow of hot, diffuse  gas
with an opening angle that increases over time,  which is often 
called the free wind.  Within this region, gas densities are low,
$V_t$ and temperture are at their highest values, and the turbulent
length scale, $L$, increases to values approaching a kpc as the gas 
rapidly expands above and below the disk.

As discussed in Strickland \etal (2004), there is considerable
observational evidence supporting the idea of superwinds
being driven by the collective  input of all the  massive stars near the  central
regions of the galaxy, rather than by the Rayleigh-Taylor break-up of
individual superbubbles.  Observed superwind pressure profiles, for
example, demonstrate that mass and energy are ejected relatively
smoothly over large regions (Heckman \etal 1990; Lehnert \& Heckman
1996), and  the edges of well-resolved outflows match up well with the
edges of starbursting regions (Strickland  \& Stevens 2000; Strickland
\etal 2000).  We address the observational consequences of our results
further in \S3.3 below.

\begin{figure}
\centerline{\includegraphics[width=4.0in]{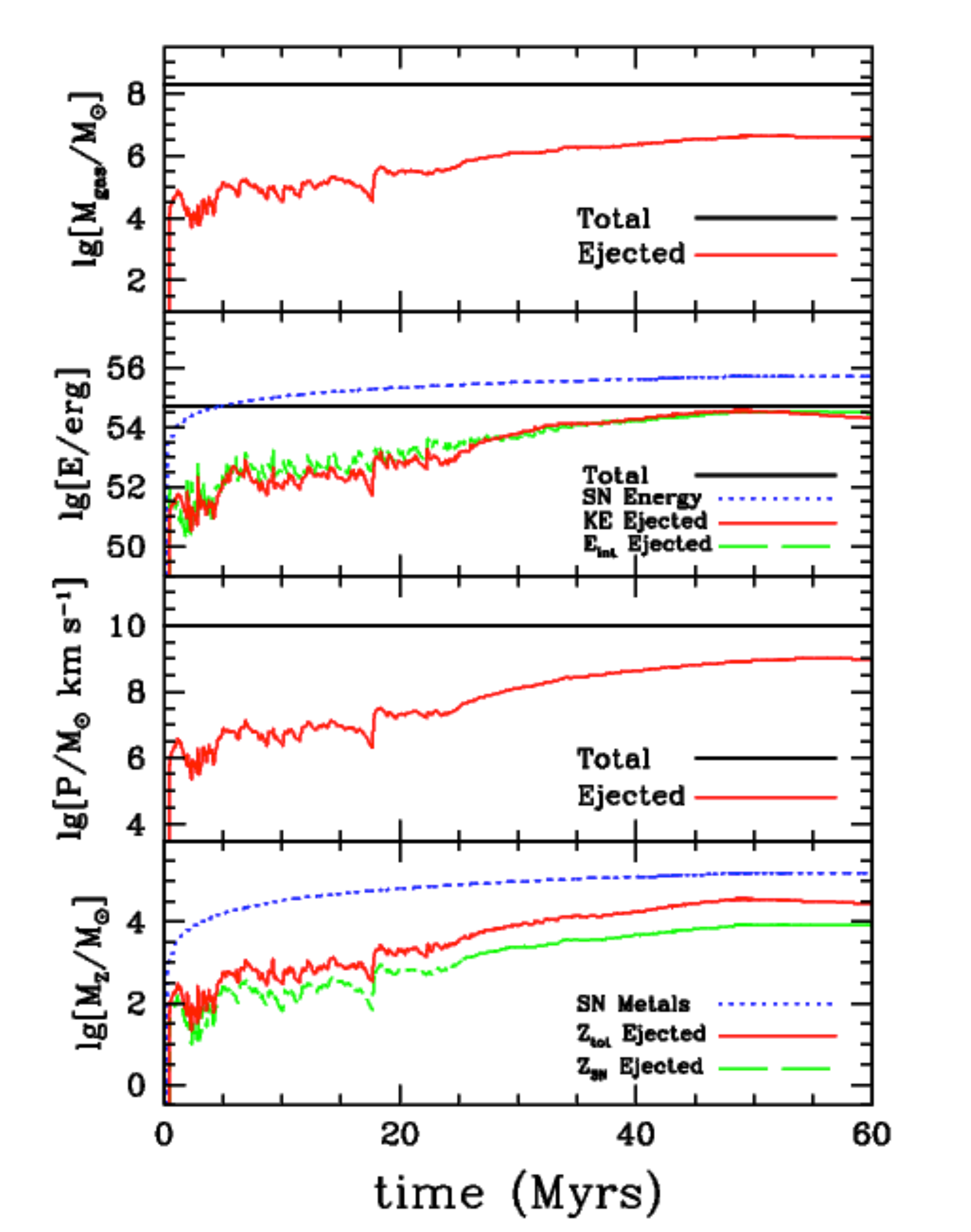}}
\caption{Evolution of ejected fractions in our fiducial simulation
  (7-4D).  {\em Top Panel:} Evolution of the ejected gas mass  (solid
  line) as compared to the total gas mass of $2 \times 10^8 M_\odot$
  (thick horizontal line).  {\em Second Panel:} Evolution of the
  ejected kinetic energy (solid line) and ejected internal energy
  (dashed line) as compared to the energy input from SNe (dotted line)
  and $M_{gas} v_{\rm esc}^2/2$ (thick horizontal line).  
The small decrease in ejected KE at 55-60 Myrs is due to the
some of the fastest moving material moving out of the simulation volume.
{\em Third    Panel:} Evolution of the total ejected momentum (solid line) as
  compared to  $M_{gas} v_{\rm esc}$ (thick horizontal line).  {\em
    Bottom Panel:}  Evolution of the ejected mass of metals (solid
  line) and metals coming purely from SNe going off during
  the simulation (dashed line), versus the total mass of metals added
  to the simulation (dotted lines).}
\label{fig:evolution_fiducial}
\end{figure}

Fig.\ \ref{fig:fiducial_xy_slice_large} shows vertical contours of the 
late-time evolution of the starburst.  After blow-out, the collective outflow
remains overpressured with respect to the surrounding intergalactic
medium, even at large distances.  While the highly rarefied free wind
does eventually collect up a denser shell of intergalactic
gas during this expansion,  mass loading arises mainly from
dense clumps of interstellar material that are gradually mixed into the
diffuse gas.  Much of this material comes from the conical
shear interface between the free wind and the surrounding galaxy, but
there is also a contribution from clumps of gas being evaporated
directly in the path of the outflow.  Together these mixed, entrained
components provide most of the observational constraints on
starburst-driven winds.

Metals can escape from the galaxy either by being entrained in the wind
or by being directly ejected in the SNe remnants driving
the outflow.  To distinguish between these two contributions, in Fig,
\ref{fig:fiducial_metals_rendered} we show rendered images of log
$\rho_{\rm metals}$ separated into the component arising from the
metals initially in the ISM, and metals that arise from SNe 
occurring during the simulation.  Here we see that despite the fact that
turbulence is driving the outflows in our galaxies, the mixing  of
supernova ejecta into the ISM is minimal, and each of the two metal
components evolves completely differently.    Consistent with the blow-out 
picture of outflow generation,  the ISM
metals are swept up into dense shells of gas that remain
bound to the galaxy.   On the other hand,  SNe ejecta are only 
weakly mixed into the shells, such that the metallicity changes
very little within dense regions during the starburst, as observed in 
the HII regions in NGC 1569 (Kobulnicky \& Skillman 1997). 
Instead SNe metals are mostly found within the rarefied high-pressure
regions, and are able to  escape to large distance following bubble overlap.

To quantify the kinematics of the ejected gas  further, we plot in
Fig. \ref{fig:evolution_fiducial} the evolution of the ejected mass,
energy, momentum, and metals.  In the upper panel of this figure, we
show the gas mass ejected by the galaxy as a function of time,
compared to the total initial gas mass.    Here we define escaping gas
as material that is at least a scale height from the  midplane of the
galaxy and traveling outwards such that the component of velocity in
the direction of the gravitational acceleration exceeds the local
escape velocity.

The wind is able to effectively blow out $\approx 3 \times  10^6
\msun$  of gas from the dwarf starburst, which is comparable to the
total mass of stars formed, as observed in local starbursts (Martin
1999).  However, the ejected mass is relatively small as compared to
the total $2 \times 10^8 \msun$ gas mass of the galaxy, and thus the
majority of the ISM remains bound.    This is true even though the
total kinetic energy from SNe added to the simulation,  shown in the
second panel of Fig. \ref{fig:evolution_fiducial}, exceeds the binding
energy of the galaxy.  Rather the majority of the energy deposited
near the galaxy centre is carried away vertically by
the outflowing wind, while
most of the energy in superbubbles at larger radii decays to thermal
energy and is radiated away.  In fact, as discussed in Mac Low \&
Ferrara (1999), ``blow-away'' of the ISM only occurs when the lateral
walls of the central outflow are accelerated so quickly that at the
end of the blow-out phase they are moving with enough momentum to
sweep out the rest of the ISM gas radially.    This requires
significantly more energy than deposited here,  although the condition
for blow-away is far less restrictive for relatively round and puffed
up galaxies such as ours than it is for larger and less turbulent
disks (Mac Low \& Ferrara 1999).

In the third panel of Fig.~\ref{fig:evolution_fiducial} we compare the
total momentum of the ejected gas, which is  at least an order of
magnitude smaller than the total galaxy gas mass times the escape
velocity.   At late times $KE_{\rm Ejected} / P_{\rm ejected} \approx
200$ km/s indicating that the majority of the outflow escapes at a few
times the escape velocity, although again most of this mass is in the
form of the largely invisible free wind rather than the much better
observed entrained material.  

In the bottom panel of this
figure we compare the metal mass produced during the starburst to the
total ejected metal mass and the ejected metal mass coming from SNe
that occur during the simulation.
As seen in Fig.\ \ref{fig:fiducial_metals_rendered} the majority of
metals ejected from the galaxy come from the supernovae that are
driving the wind itself, while only a small fraction comes from metals
already contained within the ISM.   However,  most of
the metals produced by the starburst are  retained by the galaxy,
confined to hot regions that mix into the dense ISM on time scales much
longer than the starburst itself.

\subsection{Parameter Dependencies}

\subsubsection{Wind Efficiency}

\begin{figure*}
\centerline{\includegraphics[width=7.0in]{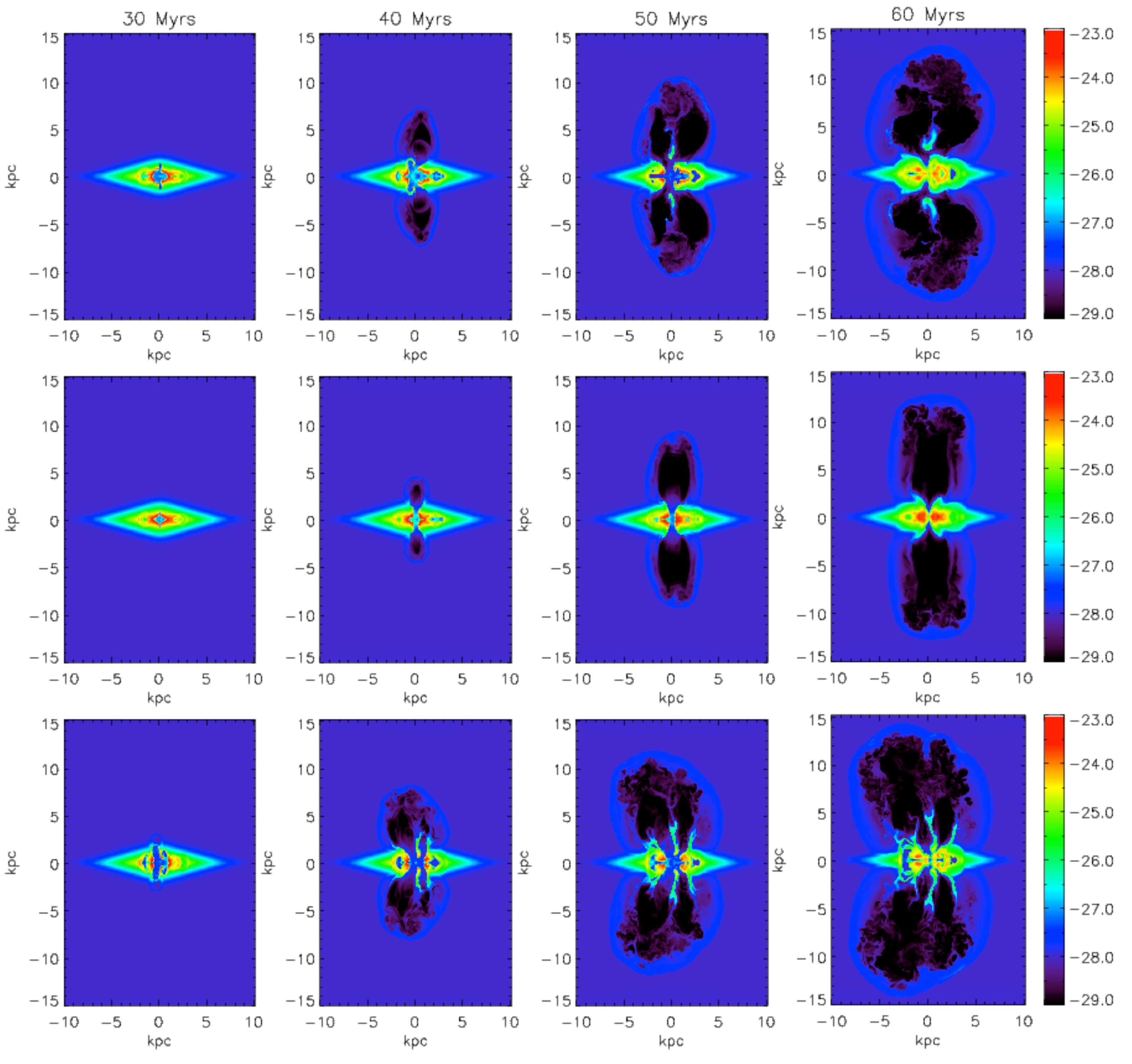}}
\caption{Vertical slices of $\log \rho$ from  $\rho =
  10^{-29}$ to $10^{-23}$ g cm$^{-3}$, through the central $20 \times
  30$ kpc in our comparison simulations (4-4D, top, 2.5-4D middle,  and
  4-4N bottom). From left to right $t$ = 30, 40, 50, \& 60 Myrs,
  respectively.}
\label{fig:comparison_xy_slice}
\end{figure*}

\begin{figure*}
\centerline{\includegraphics[width=6.0in]{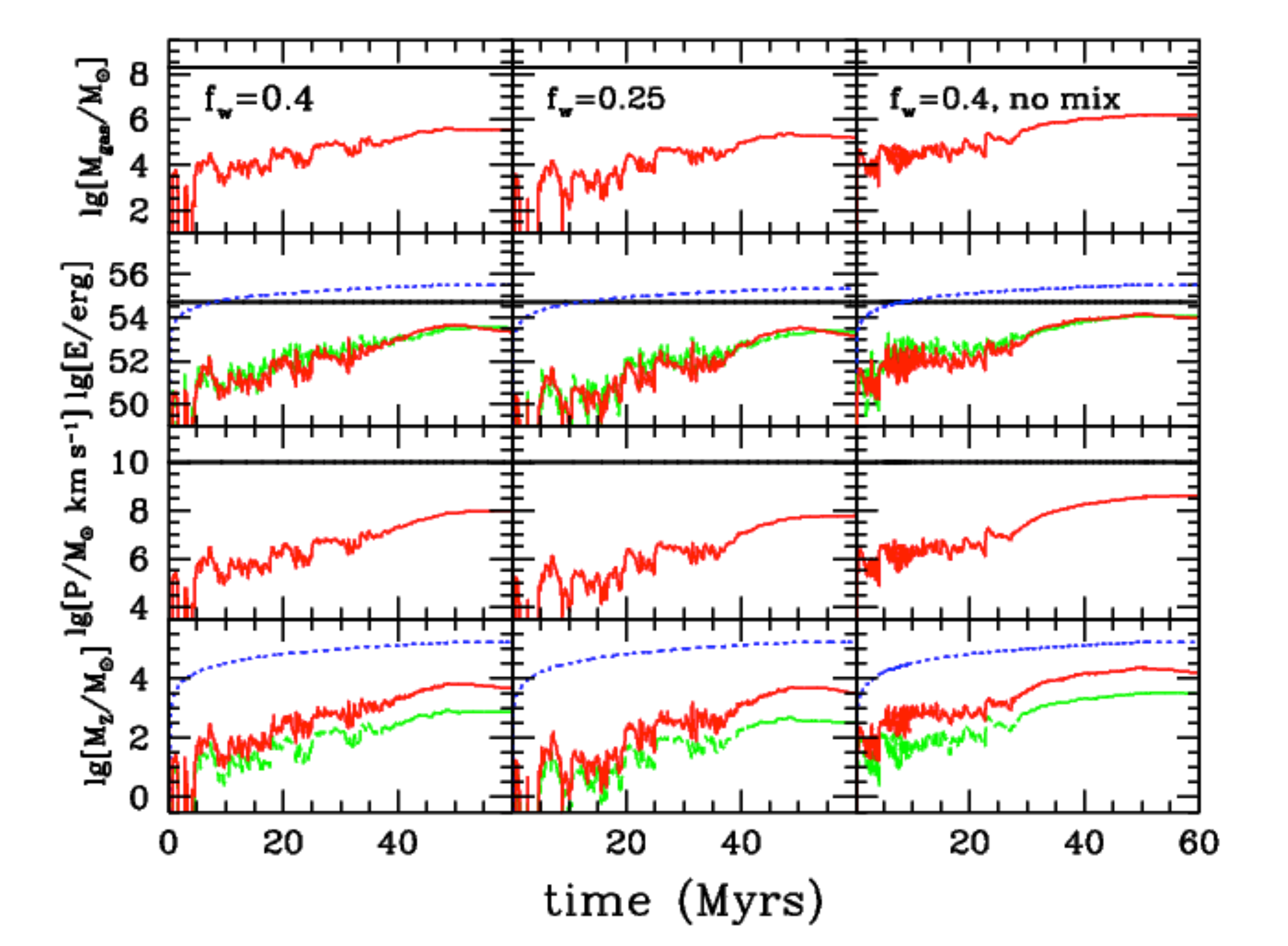}}
\caption{Evolution of the gas mass  (top), energy (2nd row), momentum (third
  row) and metal mass (bottom) for runs 4-4D (left
  column), 2.5-4D (right column), and 4-4N (left column).   Lines are as in
  Fig.\ \protect{\ref{fig:evolution_fiducial}}.}
\label{fig:evolution_compare}
\end{figure*}

In Figs.\ \ref{fig:comparison_xy_slice} and
\ref{fig:evolution_compare} we study the impact of reducing the
efficiency with which SNe energy is converted into turbulent energy
($f_{\rm w}$) and the efficiency of turbulent mixing ($C_\mu$).  In
the top two rows of Fig. \ref{fig:comparison_xy_slice} we show
vertical slices of density for runs 4-4D and 2.5-4D, for which $f_{\rm
  w}$ has been reduced to 0.4 and 0.25 respectively.   Here we see that
varying $f_{\rm w}$ leads to drastic differences in
outflow strengths and morphologies.  The galaxy in our fiducial
(7-4D) run achieves blow-out within 30 Myrs from the start of the
simulation, rapidly expanding to fill most of the simulation volume by
50 Myrs, but reducing $f_{\rm w}$ delays blow-out to $\approx$ 40-50 Myrs in the
$f_{\rm w}=0.4$ and $f_{\rm w} = 0.25$ cases, and
substantially reduces the volume enriched by the outflowing gas.
This can be understood in the context of a collective wind that arises
from overlapping superbubbles.
From eq.\ (\ref{eq:overlap}), we see directly that at a fixed radius,
overlap occurs much  later in models with low $f_{\rm w}$ as the
individual superbubbles are much more compact.  Note  also that the
opening angle of the outflow is reduced as $f_{\rm w}$ decreases, and
the central collective outflow is only able to  punch its way through
a smaller region of the ISM.   Again this is consistent with
eq.\ (\ref{eq:overlap}) which shows that for a given time,  it is only
the most central regions that can achieve superbubble overlap,
followed by  ``blow-out'' of the gas almost vertically.  Thus in the
$f_{\rm w}=0.25$ case, not only does a small fraction of the hot gas
escape from the galaxy, this is concentrated into a  plume
traveling almost directly perpendicular to the disk.

In the lower panel of this figure, we examine the effect of reducing
turbulent mixing by setting $f_{\rm w}=0.4$ and $C_\mu=0$ in a case we
label as 4-4N, where N indicates that no subgrid diffusive mixing has
been modeled. In this run, the mixing between bubble interiors and the
shells is severely reduced, which in turn reduces radiative losses,
and results in  somewhat larger superbubbles.  In general, this run
behaves similarly to our fiducial case in which we assume more
efficient SNe driving of turbulence, but mix a significant fraction of
the energy into dense, rapidly radiating regions.

In Fig.\ \ref{fig:evolution_compare}, we plot the ejected gas mass,
energy, momentum, and metal mass for each of the runs shown in
Fig.\ \ref{fig:comparison_xy_slice}.  As apparent from the density
slices, the ejected mass and momentum depend extremely sensitively on
$f_{\rm w}$.    After 60 Myr the galaxy with $f_{\rm w}=0.4$ has
ejected $\approx  10 \%$ as much mass as in the fiducial $f_{\rm w} =
0.7$ case, while the  $f_{\rm w} =0.25$ galaxy has ejected only
$\approx 3 \%.$ The differences in ejected energy  and momentum
between the runs are even more dramatic.   The energy ejected in the
$f_{\rm w} = 0.25$ run, for example, is less than $1\%$ that of the
fiducial, $f_{\rm w} = 0.7$ run.   This large difference also
translates into a large difference in the ejected metal fraction,
which is almost negligible in the $f_{\rm w} = 0.25$ case.   Finally,
in the no mixing case with $f_{\rm w} = 0.4,$  ejection of mass,
energy, momentum, and metals are all significantly enhanced, reaching
values similar those in  our fiducial run.

\subsubsection{Resolution}

\begin{figure*}
\centerline{\includegraphics[width=7.0in]{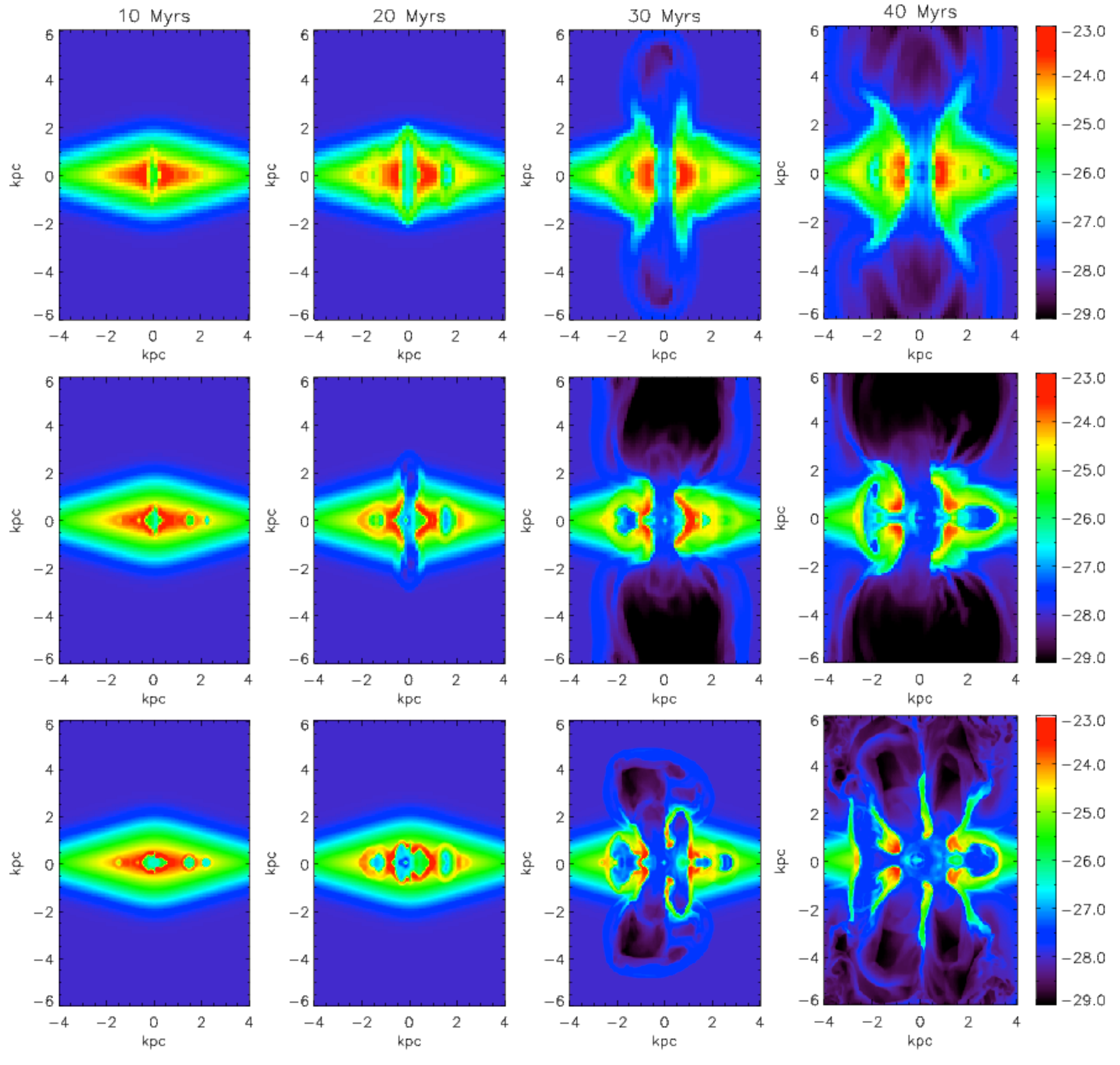}}
\caption{Vertical slices through the central $8 \times 12$ kpc in 
  simulations with varying resolutions at four representative times of
  10, 20, 30, \& 40 Myrs (arranged from left to right in each row).
  The lowest resolution run, 7-2D, is shown in the top row, the
  intermediate resolution run, 7-3D, in shown in the centre row, and
  the highest resolution, fiducial run, 7-4D is shown in the bottom
  row.  All panels give  contours of $\log \rho$ from $\rho =
  10^{-29}$ to $10^{-23}$ g cm$^{-3}$. }
\label{fig:compare_xy_slice_res}
\end{figure*}

\begin{figure*}
\centerline{\includegraphics[width=6.0in]{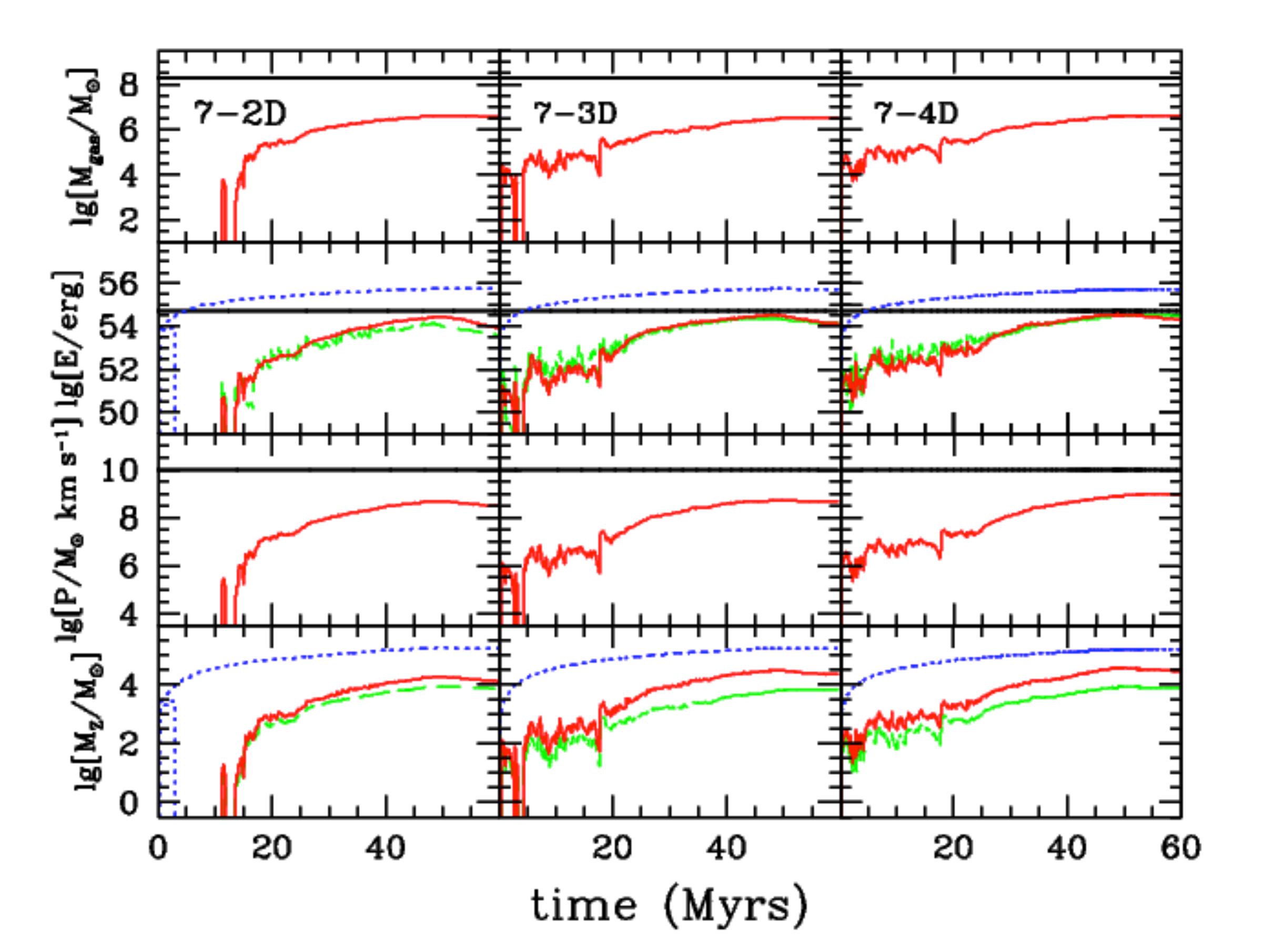}}
\caption{Comparison of the evolution obtained at varying maximum levels
  of refinement .  As in Fig.\ \protect{\ref{fig:evolution_fiducial}}, from
  top to bottom the panels show the evolution of the gas mass
  (top), energy (2nd row), momentum (third row), and metal mass
  (bottom).  From left to right the
  models correspond to 1 levels of refinement beyond the base grid
  (7-2D) yielding  156 pc resolution, 2 levels of refinement beyond
  the base grid (7-3D; 78 pc resolution), and the fiducial case
  (7-4D; 39 pc resolution).  Lines are as in Figs.\  
  \protect{\ref{fig:evolution_fiducial}} \& 
  \protect{\ref{fig:evolution_compare}}.}
\label{fig:evolution_resolution}
\end{figure*}

Next we carried out a test of the impact of resolution on our results.
Leaving the base grid spacing fixed at 313 pc, we resimulated
our fiducial $f_{\rm w} = 0.7$  galaxy  with different maximum
levels of refinement.  In the first of these runs, labeled 7-3D, we
only allowed 2 levels of refinement above the base (level 1)  grid,
for an effective resolution of 78 pc.  In the second run, labeled
7-2D, we allowed only 1 additional level, for an
effective resolution of 156 pc.  In this very low-resolution case, the
minimum value of $r_{\rm bubble}$ was taken to be 120 pc instead of 60
pc, so that each bubble region would encompass more than a
single zone.

Vertical slices at various times through the simulation volumes from these runs 
are shown in Fig.\ \ref{fig:compare_xy_slice_res}.  Note when comparing 
these runs that the turbulent length scale can grow 
larger than the grid scale,  as shown in Figures
\ref{fig:fiducial_zy_slice} and \ref{fig:fiducial_xy_slice}, 
and thus turbulent diffusion can smooth features
on scales larger than the effective resolution of each of the runs. 
With this limitation in mind, we see that in general, both low-resolution runs display 
the same  evolution as in the fiducial run.   In all cases, a low density
cavity builds up near the centre of the galaxy, pushing its way
through the lowest-density regions until it makes its way into the
surrounding intergalactic medium.   While blow-out occurs at slightly
different times in each of the runs, it is always an abrupt
transition that is quickly followed by gas ejection out to very
large distances.  At late times, in all runs, the base of the outflow
widens and the flow becomes less collimated as bubbles overlap at larger
radii and significant mass
entrainment occurs at the interface between the free wind and the
surrounding galaxy.   However, it is only in the highest resolution
run that clouds of dense ISM are resolved within the outflow
itself.

In Fig.\ \ref{fig:evolution_resolution} we compare the evolution of
the ejected quantities as a function of resolution.   At early
times, before the initial blow-out occurs, there are notable difference
between the runs, and in general, the higher resolution cases achieve
higher ejected fractions earlier.   As the simulations
progress, however, the evolution becomes much more similar between the
runs, and by the final time of 60 Myrs, all ejected quantities are
consistent to within less than a factor of two. This is true even though the
maximum volume resolution, and hence the mass resolution, varies by a factor
of $2^6$ between these runs.   This means that our implementation of supersonic
turbulence-driven outflows is only weakly dependent on the extent
to which turbulence is directly resolved, as opposed to approximated
by subgrid modeling. Instead, the qualitative and quantitive
evolution of the starburst is largely independent of resolution.

\subsubsection{ISM Structure}

Next, we examined the effects of pre-existing ISM structure on the evolution of
the outflow. As it lies outside of the scope of this paper, we did not
attempt to model a realistic density distribution of  molecular clouds.
Rather we simply altered the
gas distribution so as to add a regular series of dense $\approx 0.5$ kpc
knots,  which interact with the outflow as it develops.    Specifically,
we adopted a perturbed density distribution within the galaxy, given by: 
\ba
\rho_{\rm perturbed}(x,y,z) =  & \rho_{\rm average}(x,y,z)  [1 - 0.8 \times \qquad \qquad \qquad \nonumber\\
  & \cos(\pi x /\lambda)^2 \cos(\pi y/ \lambda)^2]^{2 \cos(\pi z /\lambda)}, 
\ea 
where $\lambda = 0.5$ kpc and $\rho_{\rm  average}$ is the smooth distribution 
given by eq.\ (\ref{eq:dens_exp_soft}).  At the same
 time we rescaled the
temperature and turbulent kinetic energy per unit mass throughout the
galaxy as $[ \rho_{\rm perturbed}/\rho_{\rm average}]^{-1}$ such that
the overall pressure profile remained the same as in the fiducial run,
with each of the dense clouds in pressure  equilibrium with its
surroundings. All other parameters for this run were identical
to the fiducial 7-4D run, and we refer to it as 7-DC, where the C
indicates the presence of a clumpy ISM.

\begin{figure*}
\centerline{\includegraphics[width=7.0in]{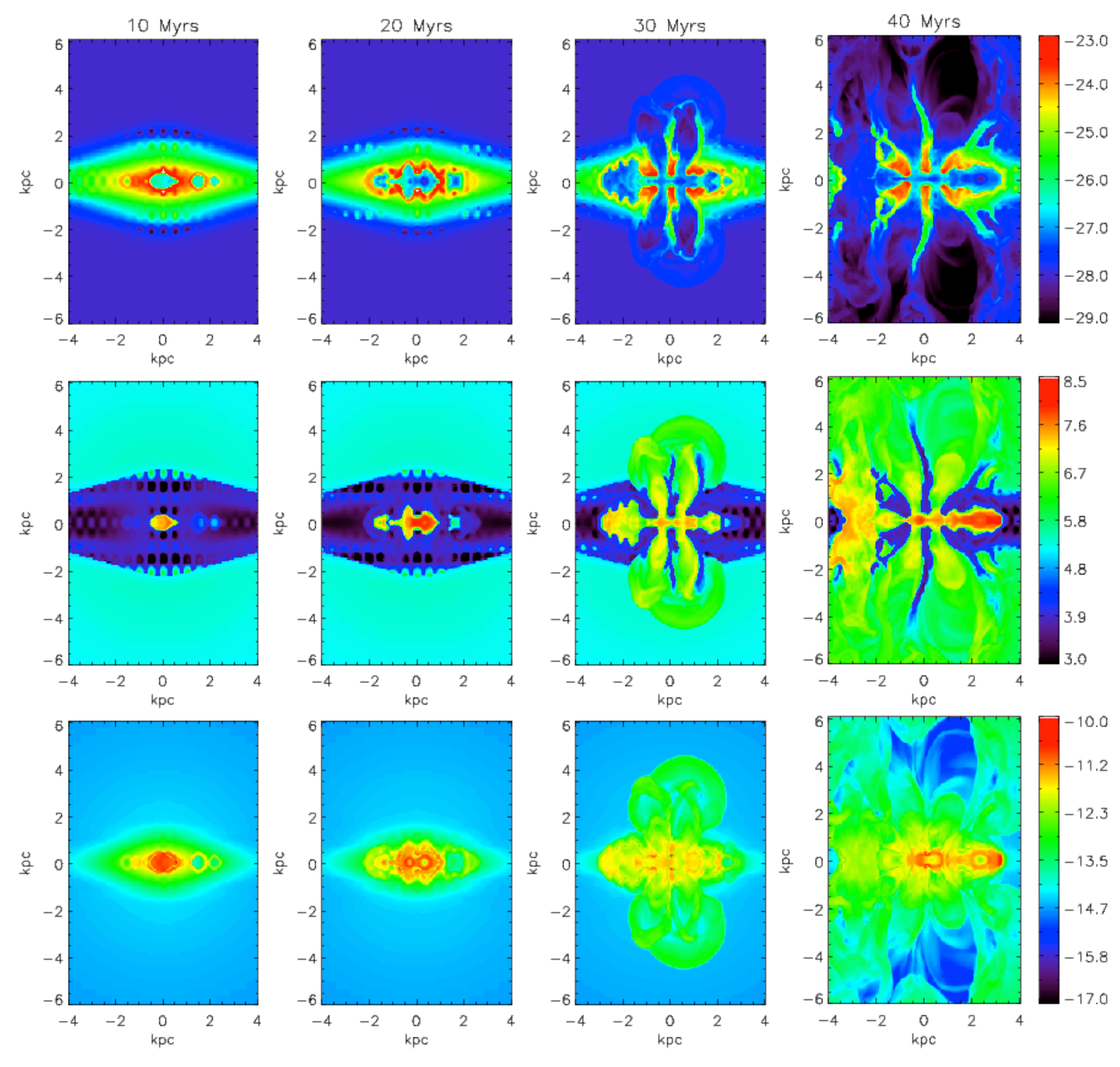}}
\caption{Vertical slices through the central $8 \times 12$ kpc in our
  perturbed simulations (7-4DC) at four representative
times of 10, 20, 30, \& 40 Myrs (arranged from left to right in each
row).
{\em Top Row:} Contours of $\log \rho$ from $\rho = 10^{-29}$ to $10^{-23}$ g cm$^{-3}$.  
{\em Centre Row:} Contours of $\log T$ from $T = 10^{3}$ to $10^{8.5}$ K.
{\em Bottom Row:} Contours of $\log p$ from $p = 10^{-17}$ ergs cm$^{-3}$ to $10^{-10}$ ergs cm$^{-3}$.}
\label{fig:lumpy_xy_slice}
\end{figure*}

In Fig.\ \ref{fig:lumpy_xy_slice} we show vertical slices of the
density, temperature, and pressure from this  simulation at several
representative times.  These illustrate the strong tendency for the
outflow to avoid dense pockets of gas.  Again, this is both
because at hig densities radiative cooling, which goes as $\rho^2,$ is  much more
efficient and because the time scale for
turbulence to decay to thermal energy is much shorter.

As in the fiducial run, superbubbles in the 7-4DC run begin to
overlap  into a large rarefied  region near the centre of the galaxy,
but the asymmetry in this run is much stronger than in the fiducial case.
Here, the gas moves primarily laterally in the
plane of the disk, so as to avoid  passing directly through the  pair 
dense clumps located directly above and below the centre.   This is an
extreme example of the cooling  instability, which preserves
dense clumps even in the direct path of the hot, high-pressure gas.
The hot region pushes its way around these clumps, forming two
distinct chimneys of hot material that punch out into the
intergalactic medium on either side of the central axis.   The free
wind then streams outwards on both sides of dense regions, enveloping
this cold gas and eventually entraining it into the outflow.  Only at
late times  does the gas within the clumps eventually begin to
move out of the galaxy, as it gradually shears and mixes into the wind
much like the ISM along the edges of the outflow.
 
Note that this evolution is markedly different than what would occur if
SN heating were modeled purely as thermal energy input and
radiative cooling were neglected.
In this case, the shocks from the developing outflow would
raise the pressure in the cold clouds, causing them to expand
and smoothing out the overall density distribution.  A tendency to
preserve a clumpy medium near the base of the wind, as observed
in NGC 1569 (Hunter, Hawley \& Gallagher 1993; Tomita, Ohta \&
Saito 1994; Heckman \etal 1995; Martin 1998; Westmoquette, Smith, \&
Gallagher 2008), is one of the hallmarks of outflows driven by
supersonic turbulence.

\subsection{Observational Consequences}

Finally, we carried out a preliminary comparison of our models with
observations of NGC 1569.   Here we focus on H$\alpha$ imaging as an
optical tracer of warm, dense, ionized gas and X-ray imaging
as a tracer of the hot, rarefied gas.  


\begin{figure*}
\centerline{\includegraphics[width=7.0in]{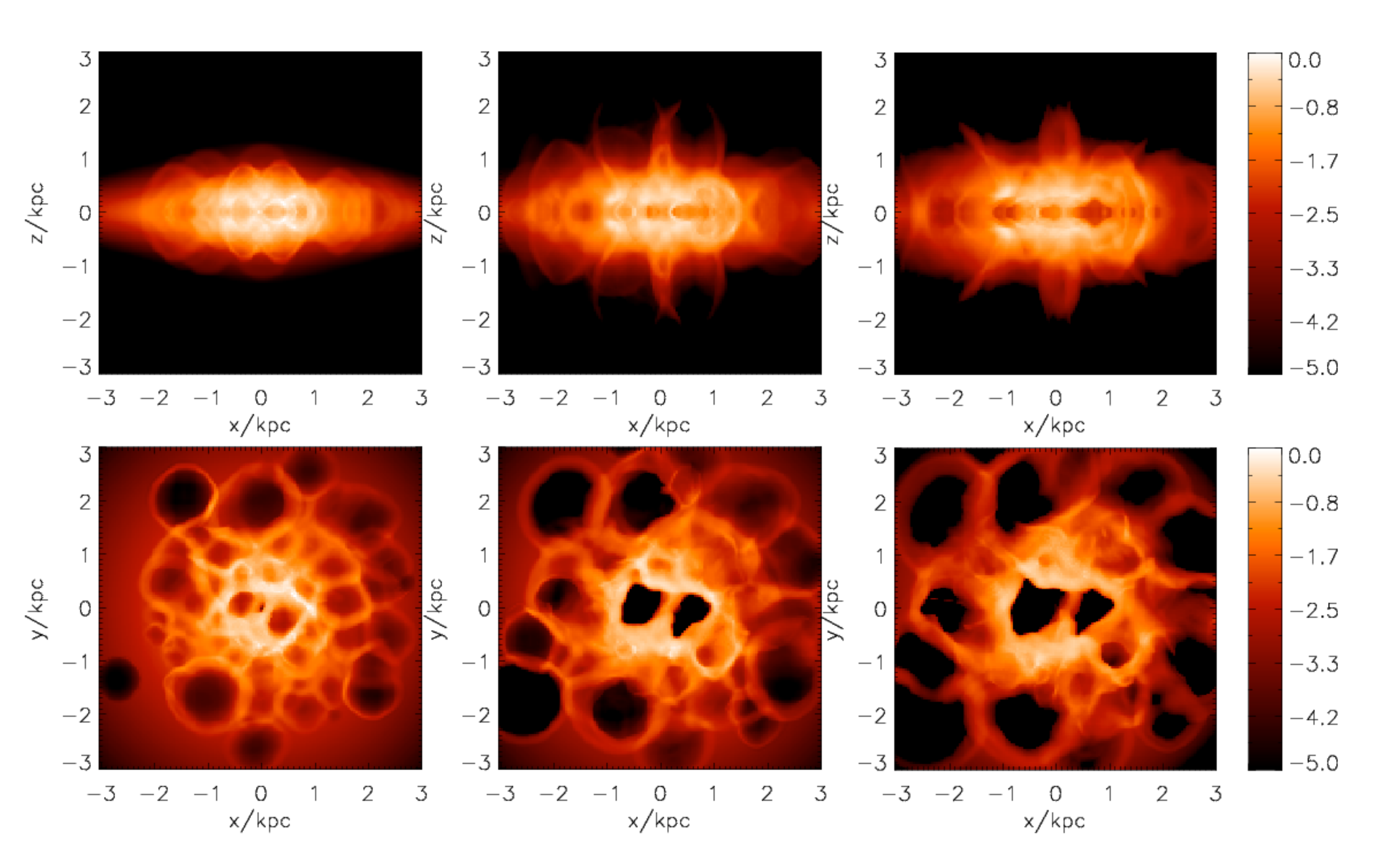}}
\caption{H$_{\alpha}$ luminosity (arbitrary scale) of our
  fiducial simulations (7-4D) at three representative
times of 20, 30, \& 40 Myrs (arranged from left to right in each row).
{\em Top Row:} Central $6 \times 6$ kpc$^2$, projected in the $z$ direction.
{\em Bottom Row:} Central $6 \times 6$ kpc$^2$, projected in the $y$ direction.}
\label{fig:halpha}
\end{figure*}

\subsubsection{H${\alpha}$ Emission}

As a tracer of the shocked and ionized interstellar medium we computed 
the H${\alpha}$ emissivity of the gas, j$_{H\alpha}$, in each grid zone 
throughout the simulation.  Note that for reasons of scope
we neglect photoionization for this calculation, although this can be very important
source in the presence of large numbers of O and B stars as would naturally occur during
a starbursts.   For this reason the images below
should be considered as only tracing the rough morphology of the ionized
gas, while more detailed studies,
such as comparisons between the velocity structure of $H{\alpha}$ emitting
gas and the coldest outflowing gas as measured though NaI absorption
(\eg Heckman \etal 2000; Fujita \etal 2009), will require more complete calculations
including radiative transfer effects.

With these limitations in mind $_{H\alpha}$ can be calculated 
as function of temperature and density from the $H\beta$ emissivity of the gas
from the theoretical fit (Ferland 1980) 
\be 
\frac{4\pi j_{H\beta}}{n_{\rm e} n_{\rm p}}=\left\{ \begin{array}{l l}
  2.53\times 10^{-22}\, T_e^{-0.833} {\rm ergs \, s}^{-1} & \mbox{for $T_e \leq 2.6\times 10^4 K$}\\
  1.12\times 10^{-20}\, T_e^{-1.2} {\rm ergs \, s}^{-1}  & \mbox{for $T_e > 2.6\times 10^4 K$},\\
\end{array} \right.
\label{eq:ferland}
\ee 
combined with an assumed Balmer decrement, j$_{H{\alpha}}$/j$_{H{\beta}}.$
For simplicity, we fixed  j$_{H{\alpha}}$/j$_{H{\beta}}= 2.9,$ ignoring 
the small  $\approx 10\%$ changes that occur over 
the range of temperatures encountered in starbursts.
Again neglecting any photoionizing background and assuming local thermodynamic equilibrium, we
calculated $n_e = n_p$ directly from the Saha equation as
\be
\frac{n_e n_p}{n_h} = \left( \frac{2 \pi m_e k T}{h} \right)^{3/2} \exp \left(\frac{-13.6 {\rm eV}}{k T} \right),
\ee
where $n_h$ is the neutral hydrogen density, $m_e$ is the mass of the
electron, and $h$ is Planck's constant. 
To account for unresolved gas inhomogeneities we assummed this
emission was enhanced by a  factor of $ 1 + 0.25 (V_t/c_s)^2,$ as in 
eq.\ (\ref{eq:ecool}).
Finally we projected the
total H${\alpha}$ emissivity in the $z$ and and $x$ directions, to
produce vertical and horizontal surface brightness maps of our
fiducial (7-4D) starbursting galaxy.

Plots of the logarithm of the surface brightness at times of 20, 30, and 40 Myrs are
given in Fig.\ \ref{fig:halpha}.  As in NGC 1569, the simulated vertical
H$\alpha$ profiles shown in the upper panels of this figure
exhibit a complex and chaotic structure,
which is is brightest near the plane of the galaxy.   Also as seen in
NGC 1569, which is observed almost edge-on, our images display
bubbles and loops of strong H$\alpha$ emission,  which correspond to
the shells of material swept up by the superbubbles generated by each
OB association (\eg Hunter, Hawley, \& Gallagher 1993; Martin 1998;
Westmoquette, Smith, \& Gallagher 2008).  

Furthermore, and especially at late times, outgoing filaments of
heated gas are apparent near  the central axis (Tomita, Ohta \& Saito
1994; Heckman \etal 1995; Martin 1998).  This is also consistent with
observations, and  a comparison of these plots with the density
contours in Fig.\ \ref{fig:fiducial_xy_slice} shows that these
features arise largely  from ISM material that is being entrained by
the hot wind, either because the dense gas is on the edges of the blow-out
region, or because it is contained in cold clumps that are left
directly within the path of the collective central outflow.
Again, these  clumps form naturally in our models from the
material swept up  between superbubbles, and they are enhanced  by the
cooling instability and the tendency for supersonic turbulent gas to avoid density
perturbations rather than disrupt them.

In the lower panels of Fig.\ \ref{fig:halpha} we show 
horizontal, $z$-projected views of the H$\alpha$ distribution in our
simulations.   While these cannot be directly compared to
observations of NGC 1569, they nevertheless serve to illustrate the cellular
nature of the warm, ionized gas, which becomes more and more
concentrated into the shells of superbubbles as time progresses.
Note also the dark gap in the centre of the images, which
grows over time.   This is the signature of blow-out, which
allows us to peer straight through the centre of the galaxy from this
vantage point, unimpeded by the warm dense medium that previously kept
the X-ray emitting central gas confined to the central starburst.

\subsubsection{X-ray Emission}

\begin{figure*}
\centerline{\includegraphics[width=6.0in]{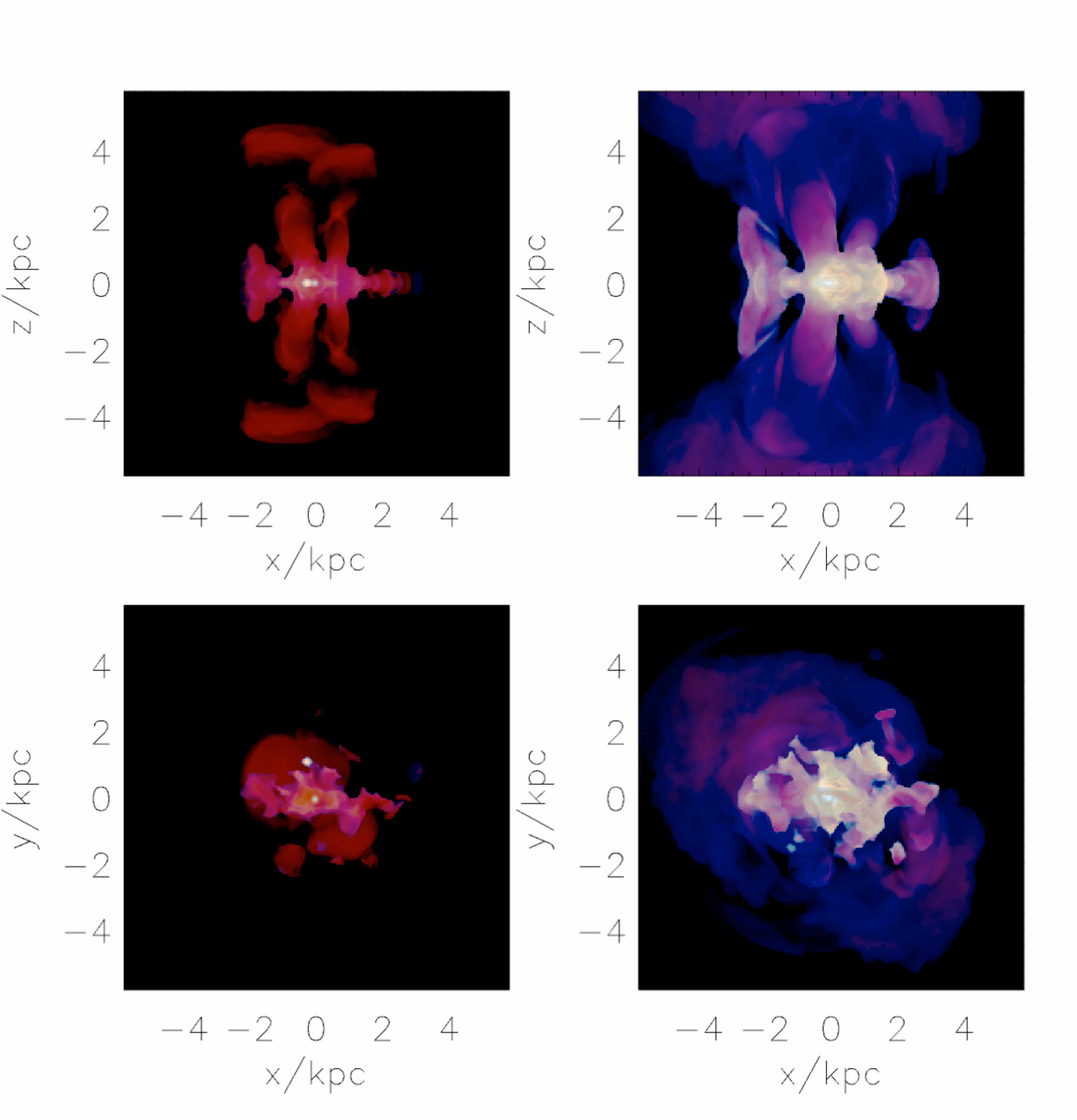}}
\caption{False colour maps of the logarithmic
  surface brightness in the  soft (0.5-2 keV; red) and hard (2-10 keV; blue) X-ray bands
  in our fiducial simulation (7-4D) at   30 (left panels) \& 40 Myrs (right panels). 
  The top row shows a projection in the $y$-direction and the bottom
  row shows a projection in the $z$-direction, spanning the central $12 \times 12 $ kpc.}
\label{fig:xray1}
\end{figure*}

Next we turn our attention to the X-ray properties of our simulated
galaxy.   Such observations provide  the most direct picture of
outflowing starbursts as they reveal diffuse emission not only from
the disk and the halo, but from the hottest and highest pressure
regions of the flow.  Martin \etal (2002), for example, have taken
Chandra observations of NGC 1569 and found that its X-ray luminosity
is dominated by diffuse, thermal emission from the disk (0.7 keV) and
a bipolar 0.3 keV halo. After subtracting hard point sources, they
found a luminosity from the thermal component in the band of 0.3-6 keV
of $\approx  8\times 10^{38}$ erg s$^{-1}$.
Moreover, Chandra observations of galactic winds have shown that the
X-rays are often spatially-correlated with H$\alpha$ emission
(Cecil \etal 2002; Strickland \etal 2004; Grimes \etal 2005).  A
prominent example is the edge-on spiral NGC 3079 (Cecil \etal 2002),
which shows towers of intertwined H$\alpha$ filaments that line its
central outflow and emit in the X-ray, as discussed by Strickland \etal
(2002).

From a theory point of view, Suchkov \etal (1994)  computed the X-ray
emission from two-dimensional simulations of galactic superwinds that
interacted with a two-component ISM. They found that the bulk of the
soft X-ray emission originated from shocked material in the disk
and halo, and the bulk of the hard X-ray emission arose  from the
free wind itself.  Consequently, they concluded that soft X-ray spectra
need not show abundances enhanced in metals. Strickland \& Stevens
(2000) also performed two-dimensional starburst simulations, computed
the X-ray emission, and  found that most of the soft X-rays come from
shock-heated ambient gas and from the interface between the hot gas
and the ISM. However, the symmetry of these simulations did not allow
them to form filamentary structures to the same degree as in
three-dimensional models, and this limited their predictions.

For this study, we computed the soft (0.5-2 keV) and hard (2-10 keV)
X-ray emissivities of our fiducial simulation using the ATOMDB
code\footnote{http://cxc.harvard.edu/atomdb/}, which includes the
Astrophysical Plasma Emission Database (APED) and the spectral models
output from the Astrophysical Plasma Emission Code (APEC). The APED
files contain information such as wavelengths, radiative transition
rates, and electron collisional excitation rate coefficients, and APEC
uses these data to calculate  model spectra, for optically-thin
plasmas in collisional ionization equilibrium. 
We also included subgrid enhancement as in  eq.\ (\ref{eq:ecool})
and projected these
metal-dependent emissivities in each grid cell to give the  surface
brightness maps shown in Fig.~\ref{fig:xray1},
which shows the distribution at 30 and 40 Myrs.
In these composite
colour images,  the red colours correspond to the soft band, the blue
colours correspond to the hard band, and both  colour scales cover six
orders of magnitude.

This figure shows that most of the soft X-ray emission comes from the
the winds that are blown out of the disk. Also, we find towers of
X-ray emission that rise more than 4 kpc above the disk and follow the
channels through which the hot wind escapes
(cf.\ Fig.~\ref{fig:fiducial_xy_slice}). The hard X-rays, on the other
hand, come mainly from the starbursting bubbles, most of which are in
the centre of the galaxy.
Comparing Figs.~\ref{fig:halpha} and \ref{fig:xray1}, it appears
that the sites of X-ray emission are correlated, similar to the
observations of NGC 3079. This is because the outflow is directly
responsible for accelerating filaments of cold gas out of the disk.

The X-ray emission is very temperature dependent and thus the emission
changes over time. At 30 Myrs the
soft X-ray emission dominates the emission from the wind and the
regions that are emitting hard X-rays are confined to the starbursting
bubbles. At this time, the overall X-ray luminosity is  $\approx 6\times
10^{38}$ erg s$^{-1}$ in the 0.5-2 keV band and $\approx  3\times
10^{35}$ erg s$^{-1}$ in the 2-10 keV band. The mean X-ray weighted
temperature of the soft X-ray emitting gas is $1.6\times 10^6$ K and
of the hard band $7.6\times 10^7$ K. The soft X-ray emission is so
luminous because most of the emission measure occurs at temperatures 
at which emission is strong. Thus the X-ray luminosity is very close
to the value measured for NGC 1569.

At 40 Myrs, the wind has moved much of the gas to higher temperatures
with the effect that the emissivity in the soft band drops, while the
emissivity in the hard band remains roughly constant. The overall X-ray luminosity
is  $\approx 9\times 10^{36}$ erg s$^{-1}$ in the 0.5-2 keV band and
$\approx  4.5\times 10^{35}$ erg s$^{-1}$ in the 2-10 keV band. The
mean X-ray weighted temperature of the soft X-ray emitting gas is
$5.6\times 10^6$ K and of the hard band $3.7\times 10^7$ K.  At 40
Myrs, emission from the free wind can been seen in the hard X-ray
band, moving out to large distances and escaping from the potential
well of the galaxy (seen as blue patches in Fig. \ref{fig:xray1}). 

However, one should note that we do not include a hot X-ray emitting
halo, and have calculated only the thermal emission from the shocked
disk and the supernovae.  Especially in the hard X-ray band
non-thermal emission becomes important. Real spectra show a
combination of thermal emission and non-thermal
components due to X-ray binaries, young supernova
remnants, low-luminosity AGN, and Compton scattering of relativistic
electrons by the ambient far-IR and CMB radiation fields (see
\eg Persic \& Rephaeli 2002). Therefore, a detailed comparison to
observations is not straightfoward, especially in the hard band, and 
for reasons of scope we postpone this to a future publication.

Finally, we note that a major driver of future X-ray instruments such as 
the X-ray Microcalorimeter System on board of the International X-ray
Observatory (IXO)\footnote{http://ixo.gsfc.nasa.gov/} is to take high-resolution
spectra of galactic winds in order to measure their velocity, temperature and abundance
structure.  While models such as ours that are able to capture turbulent broadening
of lines will be essential for this mission,  
we do not attempt to show spectra in this study as a thorough
simulation of the X-ray emission would need to  include a number of
additional important processes such as absorption. Again, this is left for
future work.

\section{Summary and Conclusions}

Starburst-driven outflows play a key role in structure formation,
impacting issues ranging from the gas and metallicity evolution
of  galaxies to the chemical and thermal history of the intergalactic
medium. While much observational and theoretical progress had been made
in  understanding these objects, formidable challenges
remain.  Many local outflowing starbursts have been
observed in great detail, but 
the phase that contains 90\% of the ejected energy and metals has gone largely undetected.
Many theoretical studies of galaxy outflows have
been conducted, but these have been faced with large uncertainties
arising from rapid radiative cooling and a complex turbulent gas
distribution that contains structures over a wide range of scales.  In
fact, the medium within starbursting galaxies is disturbed so
strongly, and the cooling times within the gas are so short,
that the turbulent velocities far exceed the thermal velocities.    

Here we have explored a completely new theoretical approach that
addresses these issues  by tracking not only the thermal and bulk
velocities of the gas, but also its turbulent velocities, pressures,
and length scales.  By adding an intermediate class of gas motions
that operates on scales much larger than the particle mean free path,
but much smaller than the resolved motions in the simulation, we are
able to carry out starburst simulations that overcome many of
the problems seen in previous models. In particular, our approach
allows us for the first time to model starbursting galaxies such as NGC
1569 without imposing a two-phase medium by hand, but still 
including realistic radiative cooling throughout the simulation.

The resulting three-dimensional AMR simulations reproduce a number of
key observational features of nearby starbursts,  some of which have
been previously captured in hydrodynamic simulations without gas cooling 
and some of which are unique to simulations that include supersonic turbulence.
Thus, in accord with previous studies, we find that:

\begin{itemize}

\item With realistic choices of star formation rates and energy input,
  our simulations lead  to  large, bipolar outflows that drive
  substantial amounts of gas, metals, momentum, and energy  into the
  intergalactic medium.    These exhibit a ``blow-out'' morphology, in
  which the majority of the ISM is retained, but a large fraction of
  the extremely hot SN-driven gas escapes in a diffuse and rapidly-expanding 
  ``free wind.''   The properties of these outflows are only
  very weakly dependent on simulation resolution, and they  occur without
  invoking any  additional physical mechanisms such as
  cosmic ray pressure or radiation pressure on dust.

\item Most mass entrainment from the ISM occurs in the shear
  interface between the free wind and the denser ISM medium.  Unlike
  other simulations with an initially homogeneous medium, however,
  this shearing  occurs both along the edge of the wind and from
  cool clouds directly in the path of the wind.   This interacting gas
  leads to the majority of the $H\alpha$ emission from the galaxy.

\item X-ray images of the galaxy show that most of the soft X-rays
  originate from the shocked material in the disk and from gas
  interactions between the hot winds and the dense ISM gas. 
  Regions of X-ray
  emission roughly correlate with regions of H$\alpha$ emission, but
  in hard X-ray images, weaker emission from the free wind can been
  seen directly, moving out to large distances and escaping from the
  potential well of the galaxy. 

\end{itemize}

At the same time, in contrast with previous studies, we find that:

\begin{itemize}

\item  Structures in our simulations arise primarily from the
  interaction of shells around individual OB associations, which sweep
  up thick shells of material around more rarefied pockets of hot gas.
  Unlike in simulations without ISM cooling,  these dense regions
  persist for long times due to the cooling instability and the
  tendency for turbulence to decay away quickly in dense regions of
  gas.   These effects lead to inhomogeneous structures throughout
  the starburst, which are far more important than the
  Rayleigh-Taylor instability in determining the outflow morphology.
  The result is  a complex, chaotic $H\alpha$ distribution, full of
  bubbles, loops and filaments, as observed in NGC 1569 and other outflowing 
  starbursts.  

\item  Outflows develop not from a single large superbubble, but
  rather from the collective action of series of smaller bubbles that
  overlap near the centre of the simulated galaxy,  where
  star formation occurs most vigorously.  These repeated  outbursts
  open an expanding, rarefied region near the galaxy centre, which 
  eats away at the denser exterior gas through
  turbulent mixing, rather than gathering it into a thin, fragile
  shell.  As a result, the rarefied region drills its way outwards
  almost directly vertically, following the path along which the
  minimum amount of material separates the bubble interior from the
  intergalactic medium.   

\item  Blow-out occurs when the overpressured  bubble regions from
  different OB associations overlap and push their way out into the
  intergalactic medium, rather than when the material surrounding a
  single superbubble becomes Rayleigh-Taylor unstable.  This means
  that the strength of the outflow is strongly dependent on the
  strength of SN driving.  Weaker  outflows escape the galaxy later,
  are much more collimated, and carry far less mass, momentum, and
  energy, than outflows from  stronger starbursts.
  
\end{itemize}

While each of these features are in excellent agreement with the H$\alpha$ 
and X-ray morphology of starburst galaxies,
such observations represent only a
small fraction of the many detailed multi-wavelength constraints
currently available. In particular, a large body of
emission and absorption line spectral data  can be brought to bear in
understanding the physics of starbursting galaxies with varying masses
and outflow strengths.   We expect future comparisons with these
observations to not only confirm aspects of our models, but to lead to
significant refinements.

After all, the subgrid turbulence model presented here represents only
a first pass at a complex and multifaceted problem. Our key
point, then, is not that our models are complete, but rather that they
point towards a new direction for future research. Due to
the extremely short cooling times in starbursting galaxies, it will be
ultimately impossible to accurately simulate them without tracking
the essential cascade of random velocities that takes place between the
bulk motions and the thermal scale.   A full understanding of galaxy
outflows will only be achieved when we have first understood and
simulated the evolution of supernova-driven, supersonic  turbulence.

\section*{Acknowledgments}

We are grateful to Annibale D'Ercole, Andrea Ferrara, Crystal Martin, Liubin Pan, for helpful comments and discussions,
and the anonymous referee for a carful reading of our paper that resulted in many improvements.
MB acknowledges support from the DFG grant BR 2026/3 within the Priority Programme ``Witnesses of Cosmic History.''
All simulations were conducted on the ``Saguaro'' cluster operated by the
Fulton School of Engineering at Arizona State University.
The results presented here were produced using the FLASH code, a product of the DOE
ASC/Alliances-funded Center for Astrophysical Thermonuclear Flashes at the
University of Chicago.

\end{document}